\documentclass[english]{elsarticle}

\usepackage[english]{babel}
\usepackage[lmargin={2.5cm},rmargin={2.5cm},tmargin={2cm},bmargin={2.5cm}]{geometry}
\biboptions{authoryear}
\usepackage{graphicx}
\usepackage{subfig}

\usepackage{soul}
\soulregister\cite7
\soulregister\ref7
\soulregister\Cref7
\soulregister\cref7
\soulregister\eqref7
\usepackage{amsmath}
\newlength{\LPlhbox}

\usepackage{amsthm}
\usepackage{mathtools}
\usepackage{amssymb}
\usepackage{hyperref} 
\usepackage[right]{eurosym}
\usepackage[utf8]{inputenc}
\DeclareUnicodeCharacter{00A0}{ }
\usepackage{fancybox}
\usepackage[usenames,dvipsnames,svgnames,table]{xcolor}
\usepackage{color}
\usepackage{tabularx}
\newcolumntype{L}[1]{>{\raggedright\arraybackslash}p{#1}} 
\newcolumntype{C}[1]{>{\centering\arraybackslash}p{#1}} 
\newcolumntype{R}[1]{>{\raggedleft\arraybackslash}p{#1}} 
\usepackage{hhline}

\usepackage{algpseudocode}
\usepackage{algorithmicx}
\usepackage[plain]{algorithm}

\algdef{SE}[DOWHILE]{Do}{doWhile}{\algorithmicdo}[1]{\algorithmicwhile\ #1}%

\usepackage{cleveref}

\begin{document}

\newtheorem{theorem}{Theorem}
\newtheorem{remark}[theorem]{Remark}
\newtheorem{lemma}[theorem]{Lemma}
\newtheorem{corollary}[theorem]{Corollary}
\newtheorem{algorithmus}[theorem]{Algorithm}
\newtheorem*{OptimizationProcedure}{Overview iterative refinement procedure}
\theoremstyle{definition}
\newtheorem{definition}[theorem]{Definition}
\numberwithin{equation}{section}

\addtolength{\jot}{1ex}

    \makeatletter
    \def\ps@pprintTitle{%
       \let\@oddhead\@empty
       \let\@evenhead\@empty
       \def\@oddfoot{\reset@font\hfil\thepage\hfil}
       \let\@evenfoot\@oddfoot
    }
    \makeatother
		
\begin{frontmatter}
\title{Comparison of wait time approximations in distribution networks using (R,Q)-order policies}

\author[cg]{Christopher Grob\corref{cor1}}
\ead{christopher.grob@volkswagen.de}
\author[ab]{Andreas Bley}
\ead{abley@mathematik.uni-kassel.de}
\cortext[cor1]{Corresponding author}
\address[cg]{Volkswagen AG, Wolfsburger Straße,
34219 Baunatal, Germany}
\address[ab]{Universität Kassel, Heinr.-Plett-Straße 40,
34132 Kassel, Germany}

\begin{abstract}
We compare different approximations for the wait time in distribution networks, in which all warehouses use an (R,Q)-order policy. 
Reporting on the results of extensive computational experiments, we evaluate the quality of several approximations presented in the literature. In these experiments, we used 
a simulation framework that was set-up to replicate the behavior of the material flow in a real distribution network rather than to comply with the assumptions made  in the literature for the different approximation. First, we used random demand data to analyze which approximation works best under which conditions. In a second step, we then checked if the results obtained for random data can be confirmed also for real-world demand data from our industrial partner. Eventually, we derive some guidelines which shall help practitioners to select  approximations which are suited well for their application. Still, our results recommend further testing with the actual application's data to verify if a chosen wait time approximation is indeed appropriate in a specific application setting.
\end{abstract}


\end{frontmatter}
\newpage

\section{Introduction}

Over the last decades multiple approximations for wait time in distribution networks that employ (R,Q)-order policies have been introduced in the literature. While each publication reports some numerical results and comparison, no comprehensive study analyzing experimentally which approximation works best in which situation has been published until now. With this paper we try to provide such a comparison. 

Our motivation for this study arose during the work on an algorithm to determine optimal reorder points in distribution networks with (R,Q)-inventory policy
(\cite{Grob.2018}). This generic algorithm works with a wide range of inventory models and wait time approximations. In our computational experiments we observed that the optimal reorder points depend heavily on the chosen wait time approximation and, furthermore, that the quality of each approximation heavily depends on the characteristics of the actual network. Wait time is the stochastic delays in deliveries to local warehouses caused by stock-outs at the central warehouse.
This naturally led to the question which wait time approximations works best in which situation and, if possible, to derive some simple rules and guidelines for practitioners to select a suitable approximation in a real-world application. The need for such a comparison was also confirmed by the reviewers of that paper. 

The traditional approach used in many multi-echelon optimization publications is to approximate the stochastic effective lead time by its mean, as first done in the METRIC model by \cite{Sherbrooke.1968}. \cite{Deuermeyer.1981} later applied METRIC to divergent multi-echelon systems with batch ordering.  \cite{Svoronos.1988} presented several methods to derive the exact distributions for the case of identical local warehouses and Poisson distributed demands. As these approximations are computationally rather costly, Svornos and Zipkin also proposed approximations for the mean and the variance of the wait time in
networks with Poisson distributed demands. \cite{Axsater.2003} derived wait time approximations using a normal approximation. All of these approaches are based on Little's law and therefore lead to identical average wait times for all local warehouses. In many practical applications, however, the frequencies and sizes of  customer demands as well as the order quantities of local warehouses vary widely, which makes this approach unsuitable. \cite{Andersson.2000} therefore proposed a heuristic approach, which adjusts the wait times derived from Little's law for each local warehouse based on average demand and order quantity, but which neglects the variance of the wait times.

Three more recent publications introduce approximations that provide mean and variance of the wait time individually for each local warehouse and that do not assume Little's law. \cite{Kiesmuller.2004} proposed an approximation for an $n$-level network based on a mixed Erlang distribution of the aggregate demand. 
\cite{Berling.2014} derived several methods to estimate the wait time in a 2-level network and their numerical analysis shows good results for an approximation based on a log-normal distribution. \cite{Grob.2018} proposed to approximate the aggregate lead time demand at the central warehouse by a negative Binomial distribution and then used a variant of the approximation of \cite{Kiesmuller.2004}  to estimate the wait time.

The remained of this paper is organized as follows. In \Cref{sec:model} we present our inventory model and introduce the necessary formulas to set-up and evaluate our simulations.  In \Cref{sec:waitapproxes} we review the different wait time approximations from the literature that are considered in this study. The simulation framework is described in \Cref{sec:simu}, while the experimental results are discussed in-depth in \Cref{sec:exp}. 

\section{Inventory model}\label{sec:model}

\begin{figure}[htbp] 
  \centering
     \includegraphics[width=0.5\textwidth]{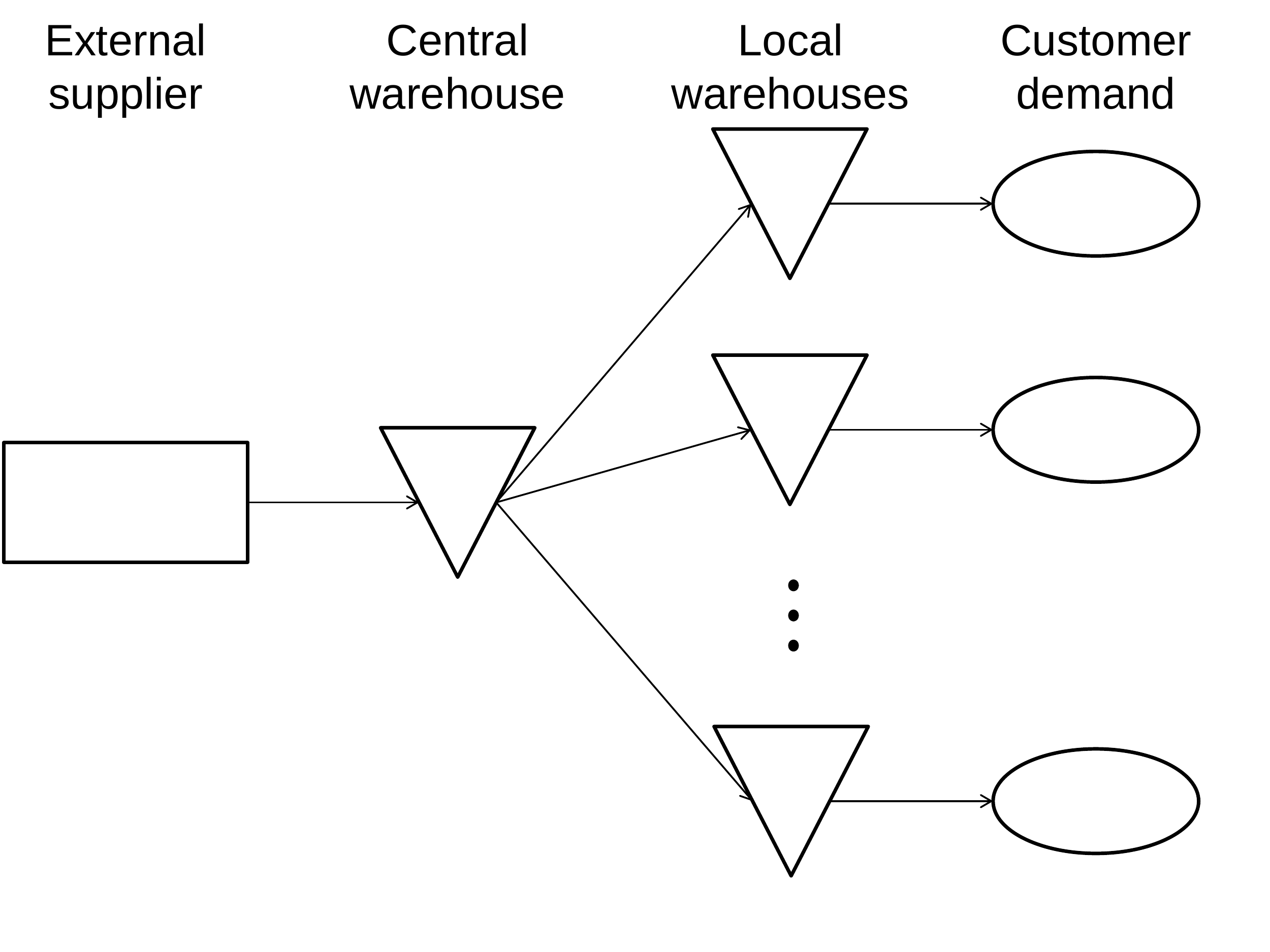} 
  \caption{Example of a 2-level distribution network}
  \label{fig:networkex}
\end{figure}

In our model, we consider a two-level distribution network. A central warehouse is supplied with parts from an external source. The external supplier has unlimited stock, but the lead time is random with known mean and variance. The central warehouse distributes the  parts further on to $n$ local warehouses, which fulfill the stochastic stationary customer demands. \Cref{fig:networkex} shows a sample network. Let the inventory position be defined as stock on hand plus outstanding orders minus backorders. The central as well as the local warehouses order according to a continuous review $(R,Q)$-inventory policy: Whenever the inventory position 
is $R$ or below $R$, a given lot size $Q$ is ordered from the predecessor in the network until the inventory position is above $R$ again.  

The local warehouses $i=1,2,...,n$ order from the central warehouse $0$, whereas the central warehouse orders from the external supplier. If sufficient stock is on hand, the order arrives a random lead time with known distribution, $L_0$ or $L_i$, respectively, later. In this paper, we assume complete deliveries. If stock is insufficient to fulfill the complete order, a backorder is placed and the (complete) order is fulfilled as soon as enough stock is available again.  Orders and backorders are served first-come-first-serve. If a stock-out occurs at the central warehouse, orders of local warehouses have to wait for an additional time until sufficient stock is available again. This additional time is modeled as a random variable, the wait time $W_i$. The effective stochastic lead time therefore is $L_i^{eff}=L_i+W_i$. $L_i$ and $W_i$ are assumed to be independent random variables. We also assume unlimited stock at the supplier. Hence, the wait time of orders of the central warehouse is $0$.

The local warehouses fulfill external customer demand. Unfulfilled customer orders are backordered until sufficient stock is on hand and are served on a first-come-first-serve basis, again.

The following table summarize our notation:

\medskip
	\begin{tabular}{ll}
		$R_i$& reorder point of warehouse $i$,\\
		$Q_i$& order quantity of warehouse $i$,\\
		$p_i$& price of the considered part at warehouse $i$,\\
		$\bar{\beta_i}$& fill rate target of warehouse $i$,\\
		$\mu_i$ & expected customer demand at warehouse $i$ during one time unit,\\
		$\sigma_i$ & standard deviation of customer demand at warehouse $i$ during one time unit,\\
		$L_i$& lead time of warehouse $i$,\\
		$W_i(R_0)$& wait time of local warehouse $i$, which depends on the central reorder point,\\
		$L_i^{eff}(R_0)$& effective lead time of local warehouse $i$, defined as $L_i+W_i(R_0)$,\\
		$D_i(t)$ & demand at the local warehouse $i$ during a time period of length $t$, and\\
		$I_i$ & inventory level of warehouse $i$.
	\end{tabular}
\medskip

We assume that customer demands arrive according to a compound Poisson process. The probability for an order of size $k$ at node $i$ is denoted by $pdf_{K_i}(k)$. 
If the order size $K_i$ follows a logarithmic distribution, the demand during a given time period (for example the lead time) is negatively binomial distributed. 

\cite{Rossetti.2011} have studied the adequacy of different distributions and distribution selection rules. They considered a single warehouse that uses $(R,Q)$-order policy and performed simulations to determine how suitable certain distributions and distribution selection rules are to model lead time demand. They conclude that the negative binomial distribution has ``excellent potential'', especially if demand variability is high. Also, our industrial partner has obtained promising results in previous studies by using the negative binomial distribution to model lead time demand. 

In order to derive a formula for the fill rates at the local warehouses, we  first need to estimate the demand during the effective lead time $D_i(L_i^{eff})$. The mean and the variance of this so-called lead time demand are
		\begin{align*}
			E[D_i(L_i^{eff})]=& \mu_iE[L_i^{eff}] \qquad\text{and}
	\\	
			\text{Var}[D_i(L_i^{eff})]=& \sigma_i^2E[L_i^{eff}]+\mu_i^2Var[L_i^{eff}] ~.
		\end{align*}
The probability that the inventory level, which is the stock on hand minus the backorders, at node $i$ is equal to $j$ can be expressed as
\begin{align*} 
	Pr(I_i^{lev}=j)=\frac{1}{Q_i}\sum_{l=R_i+1}^{R_i+Q_i}Pr(D_i(L_i^{eff})= l-j)~ .
\end{align*}
This follows from the uniformity property of the inventory position (\cite{Axsater.2006}).

If only complete deliveries are permitted, an order can only be fulfilled if the inventory level $j$ is at least as big as the order size $k$. With this approach, also referred to as \emph{order fill rate}, i.e., the fraction of orders that can be fulfilled directly from stock on hand, we obtain 
\begin{align}
	\beta_i(R_i)&=\sum_{k=1}^{R_i+Q_i}\sum_{j=k}^{R_i+Q_i} pdf_{K_i}(k) \cdot Pr(I^{lev}=j)
	\notag	\\
	&=\sum_{k=1}^{R_i+Q_i}\sum_{j=k}^{R_i+Q_i} pdf_{K_i}(k) \frac{1}{Q_i}\sum_{l=R_i+1}^{R_i+Q_i}Pr(D_i(L_i^{eff})= l-j) ~.
	\label{eq:frord}
\end{align}

In order to calculate the central depot's fill rate, we have to determine the mean and the variance of the central lead time demand first. Let $L_0$ be a random variable denoting a time period and let $pdf_{L_0}(l)$ 
denote the probability density function of $L_0$. We assume that the support of this function is positive.
Due to the uniformity property of the inventory position, $(I^{Pos}_i(t)-R_i)$ is uniformly distributed on $(0,Q_i]$. The probability of depot $i$ placing at most $k$ orders, conditioned on $L_0=l$, is
\begin{align}\label{formel_unitsordered_dis}
	\delta_i(k|L_0=l)=\sum_{x=1}^{Q_i}\frac{1}{Q_i}Pr\left(D_i(l)\le kQ_i+x-1\right) \qquad \text{for}~ k=0,1,2,\ldots~.
\end{align}
By integrating or summing over all possible times $l$, we obtain  
\begin{align}\label{formel_unitsordered_cont}
	\delta_i(k)=\int_{l=0}^\infty\left(\sum_{x=1}^{Q_i}\frac{1}{Q_i}Pr\left(D_i(l)\le kQ_i+x-1\right) \cdot pdf_{L_0}(l)\right) \mathbf{d}l
\end{align}
for a continuous distribution of the lead time  (cp.\ \cite[p.115]{Ross.2014}) or the analogous sum for a discrete distribution.
\Cref{formel_unitsordered_dis} or \eqref{formel_unitsordered_cont}, respectively, imply that the probability that exactly $k$ subbatch orders of size $Q_i$ are placed from depot $i$ is
\begin{align*}
	s_i^{ord}(k)=
	\begin{cases}
		\delta_i(0), & \text{if } k=0,\\
		\delta_i(k)-\delta_i(k-1), & \text{if } k>0,\\
		0, &\text{otherwise.}
	\end{cases}
\end{align*}
Using these terms, the mean and the variance of the lead time demand at the central warehouse can be expressed as
\begin{align}\label{formel_dlt0mu}
	E[D_0(L_0)]=&\sum_{i=1}^N\mu_iE[L_0]\qquad\text{and}
\\
\label{formel_dlt0var}
	\text{Var}[D_0(L_0)]=&\sum_{i=1}^N\left[\sum_{k=0}^\infty \left(\mu_iE[L_0]-kQ_i\right)^2 s_i^{ord}(k)\right]~.
\end{align}
Note that, if the common subbatch size $q=gcd(Q_0,\dots,Q_n)$ of all local order quantities is greater than $1$, the above calculations should be done in units of $q$, as this leads to a better the approximation of the fill rate.

Similar to \cite{Berling.2013}, we fit standard distributions to the mean and the variance of lead time demand using the following distribution selection rule: 
If $Var[D_0(L_0)]>E[D_0(L_0)]$, we use the negative binomial distribution (which only exists if this condition holds) and use \eqref{eq:frord} to calculate the fill rate. In all other cases, we propose to use the gamma distribution. This same rule is applied to choose the distributions for the local warehouses.

\section{Wait time approximations}\label{sec:waitapproxes}

In this section we present the four wait time approximations that we consider in our analysis. We have selected one METRIC-type approximation by \cite{Axsater.2003} as a representative for this class of approximations and the three approximations proposed by \cite{Berling.2014}, by  \cite{Kiesmuller.2004} and by \cite{Grob.2018}.

For the sake of simplicity, all approximations are presented only for the 2-level case. Note that the approximation by Berling and Farvid is only valid for 2-levels, while the approximation by Kiesmüller et al.\ is also applicable for the more general $n$-level case. Axsäter introduced his approximation for the $2$-level case, but he also sketched an extension to $n$-level case at the end of his paper. Also the approximation by \citeauthor{Grob.2018} has been introduced for $2$ levels and can, in principle, be extended to the $n$-level case, but the quality of the approximation is rather bad for more than $2$ levels.

\subsection{METRIC-type approximation}
For benchmark and comparison purposes a METRIC-type approximation based on Little's law is used. It was originally introduced by \cite{Sherbrooke.1968}. \cite{Deuermeyer.1981} were the first to adapt it for batch ordering systems and it was widely used in research publications since then.  Deuermeyer and Schwarz assumed identical demand and order quantities at local warehouses. If these parameters are not identical among the local warehouses, the formula only yields the average delay for all warehouses. This means that one may get significant errors if the order quantities of local warehouses differ substantially. Nevertheless, this approach is still commonly used as an approximation for the expected wait time.
Due to their simplicity, the METRIC-type approximations are probably the most widely used approach in practical applications. 

There exist some approaches which also model the variance of the average delay. \cite{Svoronos.1988} derive the variance based on an extension of Little's law for higher moments, but it is only exact if the order quantity of all local warehouses is 1. \cite{Axsater.2003} derives the variance of the average delay based on a normal approximation of lead time demand. 

We have chosen to use Axsäter's approximation as the representative for METRIC-type approximations. He derives mean and variance based on Little's law, the normal approximation of lead time demand, constant central lead time, and a continuous uniform approximation of the (discrete) central inventory position. Here, the wait times are identical for all local warehouses, therefore we drop the subscript $i$ of $W_i$. Under these assumptions we obtain
\begin{align}\label{eq:metric_mean}
	E[W]&=\frac{E[(I_0)^-]}{\sum_{i=1}^n\mu_i}=\frac{G(k(R_0))}{\sum_{i=1}^n\mu_i}\sum_{i=1}^n\sigma_iE[L_0]~\text{, and}\\
	Var[W]&=\left(\frac{E[W]}{G\left(k(R_0)\right)}\right)^2\left(1-\Phi(k(R_0))\right)-\frac{(E[W])^2k(R_0)}{G(k(R_0))}-\left(E[W]\right)^2,
\end{align}
where $G(\cdot)$ is the normal loss function and
\begin{align}
	k(R_0)=\frac{R_0+Q_0-\sum_{i=1}^n(\mu_iE[L_0])}{\sum_{i=1}^n\sigma_iE[L_0]}~.\notag
\end{align}

Axsäter compares the approximation to the exact solution for a problem instance with very low demand, as the exact solution is computable in this case. He concludes that the deviations of the approximation from the exact solution would be acceptable in most practical situations.

\subsection{Kiesmüller et al.\ approximation}\label{sec:waitkies}
\cite{Kiesmuller.2004} derive approximations for the first two moments of the wait time. They consider a compound renewal customer demand process and use the stationary interval method of \cite{Whitt.1982} to superpose the renewal processes with mixed Erlang distributions. Their model only applies for non-negative central reorder points.
They introduce an additional random variable $O_i$ for the actual replenishment order size of warehouse $i$ and approximate the first moment as
\begin{align}\label{formel_expwaitkies}
	E[W_i]\approx \frac{E[L_0]}{Q_0}\left(E\left[\left(D_0(\hat{L}_0)+O_i-R_0\right)^+\right]-E\left[\left(D_0(\hat{L}_0)+O_i-(R_0+Q_0)\right)^+\right]\right),
\end{align}
where $\hat{L}_0$ is random variable representing the residual lifetime of $L_0$ with distribution function
\begin{align}\label{formel_reslead}
	F_{\hat{L}_0}(y)=\frac{1}{E[L_0]}\int_0^y\left(1-F_{L_0}(z)\right)dz.
\end{align}
The second moment of the wait time can be approximated by
\begin{align}\label{formel_wait2momkies}
	E[W_i^2]\approx \frac{E[L_0^2]}{Q_0}\left(E\left[\left(D_0(\tilde{L}_0)+O_i-R_0\right)^+\right]-E\left[\left(D_0(\tilde{L}_0)+O_i-(R_0+Q_0)\right)^+\right]\right),
\end{align}
where $\tilde{L}_0$ is a random variable with distribution function
\begin{align}\label{formel_reslead2}
	F_{\tilde{L}_0}(y)=\frac{2}{E[L_0^2]}\int_0^y \int_x^\infty(z-x)dF_{L_0}(z)dx.
\end{align}
For further details of the calculation we refer the reader to the original paper by \cite{Kiesmuller.2004}.

Kiesmüller et al.\ report that the performance of their approximation is excellent if order quantities are large. This applies especially for the order quantities of the local warehouses. The error of the approximation reportedly also decreases with increasing demand variability.

Unfortunately, the approximation by Kiesmuller et al.\ yields negative value for the variances in some situations, an observation that was also made by \cite{Berling.2014}. 
Furthermore, the two-moment fitting technique using the mixed Erlang distribution may lead to serious
numerical difficulties if the squared coefficient of variation is very low. This happens, for example, in situations where the order quantity of the local warehouse is very large compared to lead time demand.

\subsection{Berling and Farvid approximation}\label{sec:waitberfar}

\cite{Berling.2014} estimate mean and variance of the wait time by replacing stochastic lead time demand with a stochastic demand rate.  They offer several methods to approximate this demand rate assuming that the  central lead time is constant.

We have chosen to use method "E5" from their paper, which uses an approximation based on the lognormal distribution. Berling and Farvid obtained good results with this method in their numerical analysis. Additionally, it is easy to implement and it does not require challenging computations.

Berling and Farvid compared their methods to \cite{Kiesmuller.2004}, \cite{Axsater.2003}, \cite{Andersson.2000}, and \cite{Farvid.2014}. Their method showed superior results for the cases considered in this study. Kiesmüller et al., however, state that their approximation works best with highly variable demand and large order quantities, while Berling and Farvid use very low order quantities (compared to the demand rate) and a demand process with low variance. Thus, it is not surprising that the results obtained are in favor for Berling and Farvid's approach.

Berling and Farvid's approximation does not work if the order quantity of a local warehouse is larger than the sum of the order quantity of the central warehouse and the central reorder point. In this case, a negative expected wait time is calculated. However, this scenario is not very common. We recommend setting the wait time equal to the central lead time in such a case.

\subsection{Negative Binomal approximation}\label{waittime_negbin}
 \cite{Grob.2018} propose to approximate the aggregate lead time demand at the central warehouse by a negative Binomial distribution and then compute the first two moments of the wait time analogous to \cite{Kiesmuller.2004}. The negative Binomial distribution is a flexible distribution and suitable to model lead time demand. Also, this approach does not face numerical instabilities and negative variances, as encountered by Kiesmüller et al.

In the industrial application considered in \cite{Grob.2018} and in this paper, the order quantity is large compared to lead time demand. We therefore assume that the actual order size is always equal to the order quantity. If order quantities are too small for this assumption to hold, it is easy to incorporate the changes into the model. Adapting equations \eqref{formel_expwaitkies} and \eqref{formel_wait2momkies}, one can express the first two moments of the wait time as
\begin{align}\label{formel_expwait}
	E[W_i]\approx \frac{E[L_0]}{Q_0}\left(E\left[\left(D_0(\hat{L}_0)+Q_i-R_0\right)^+\right]-E\left[\left(D_0(\hat{L}_0)+Q_0-(R_0+Q_0)\right)^+\right]\right)
\end{align}
and
\begin{align}\label{formel_wait2mom}
	E[W_i^2]\approx \frac{E[L_0^2]}{Q_0}\left(E\left[\left(D_0(\tilde{L}_0)+Q_i-R_0\right)^+\right]-E\left[\left(D_0(\tilde{L}_0)+Q_i-(R_0+Q_0)\right)^+\right]\right),
\end{align}
where $\hat{L}_0$ and $\tilde{L}_0$ are defined as in \eqref{formel_reslead} and \eqref{formel_reslead2}.

We determine mean and variance of $D_i(\hat{L}_0)$ and $D_i(\tilde{L}_0)$ with the help of \eqref{formel_dlt0mu} and \eqref{formel_dlt0var}, replacing $L_0$ with the respective random variables. Instead of deriving the real distribution function, we then fit the mean and variance to a negative Binomial distribution. If the variance is smaller than the mean, the negative Binomial distribution is not defined. This can only happen, if local order quantities and the coefficient of variation of local demand are small. In this case, we propose using the gamma distribution instead.

For a discrete distribution with positive support, such as the negative Binomial distribution, it is easy to show that
\begin{align}\label{negbin_positiveexp}
E[(X-z)^+]=E[X]-\sum_{x=0}^zxf_X(x)-z(1-F_X(z)),~ z\ge0~.
\end{align} 
A similar equation can be derived for the gamma distribution.

Given the issues with the original Kiesmüller et al.\ wait time approximation, our motivation was to develop a more stable approach for use in practical applications. The negative binomial approximation, although inspired by Kiesmüller et al.'s approximation, does not suffer from the same instabilities and it is easier to compute. However, the negative binomial approximation has issues if $Q_i \gg Q_0$. In this case, the approximate value for the variance may be negative as well, but this case is a very  unlikely setting in real-world applications. 

\section{Set-up of simulation}\label{sec:simu}

\begin{figure}[htbp]
	\centering
		\includegraphics[width=\textwidth]{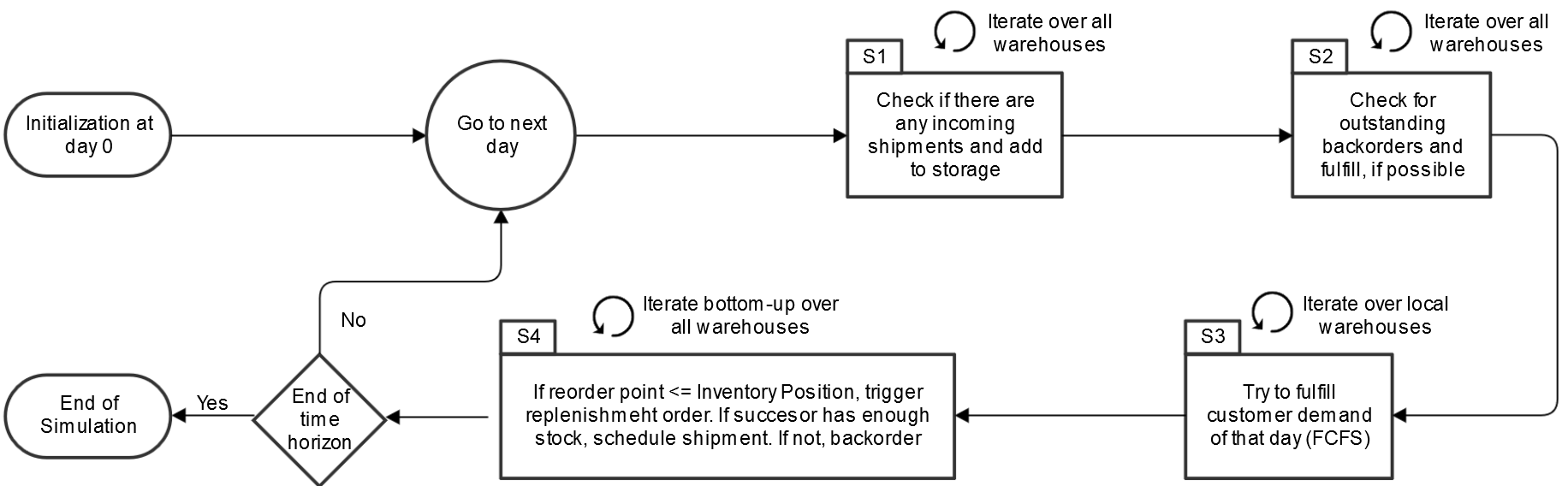}
	\caption{Schematic process diagram of the simulation}
	\label{fig_simulation_process}
\end{figure}

In this section, we describe the simulation framework that we used in our experimental study.  \Cref{fig_simulation_process} shows the process diagram of this framework, which models a divergent inventory system where all warehouses use a (R,Q)-order policy.
Our goal was to simulate the behavior of real-world warehouses as realistic as necessary, while keeping the simulation as simple as possible. Hence, we have chosen to simulate all processes based on discrete time steps and a specified processing order of receiving shipments, fulfilling demands, and placing new (back-)\,orders. In our application the discrete time step is one day, but it could also be any other time unit. The final model was derived after multiple rounds of discussion and refinement involving several practitioners from the automotive industry. We are aware that this set-up may violate some assumptions of the different wait time approximations. It can therefore also be seen as a test of robustness towards the perils of the real-world.

In the following, we describe the simulation and especially the process steps \textsc{S1} to \textsc{S4} of \Cref{fig_simulation_process} in more detail. 

For each warehouse $i$ in the network, we are given the input data shown in \Cref{tab_InputDataNeededForSimulation}.
\begin{table*}[htbp]
	\centering
		\begin{tabular}{l l}
			Input data & Notation \\ \hline
			Reorder points &$R_{i}$\\
			Order quantities & $Q_{i}$\\
			Expected transportation time & $E[T_{i}]$\\			
			Variance of transportation time & $Var[T_{i}]$
		\end{tabular}
	\caption{Warehouse input data needed for simulation}
	\label{tab_InputDataNeededForSimulation}
\end{table*}

At the start of the simulation,
the initial inventory on hand at warehouse $i$ is set to $R_{i}+1$, no shipments are on their way,  and all other variables are set to $0$. 
Then, as soon as we have incoming demands at the local warehouses, new orders are triggered and propagated throughout the network as shown in  \Cref{fig_simulation_process}.

Processing a time step, we first check in step \textbf{\textsc{S1}} for each warehouse if there are any incoming shipments and add them to the inventory on hand. 

Then, in step \textbf{\textsc{S2}}, we check for each warehouse if there are any outstanding backorders. In this step, we differentiate between local warehouses, which fulfill customer demand, and non-local warehouses, which only handle internal demand. For local warehouses, we use our inventory on hand to serve backorders first come - first serve as far as possible. For non-local warehouses, we use inventory on hand to fulfill orders of succeeding warehouses. In this case, if we fulfill an order of a successor, we also schedule the corresponding shipment to the successor. The transportation time $t$ for this shipment is drawn at random according to a given distribution. In our simulations, we used a gamma distribution, rounding the drawn random numbers to the nearest integers. The shipment is scheduled to arrive $t$ days in the future at the successor. The minimal transportation time possible is 1 day. 

Afterwards, in step \textbf{\textsc{S3}}, we iterate over all local warehouses and try to fulfill the customer demand of that day first come - first serve. Here, our simulation allows for two modes of operation: The first one is based on historical demand data, while the second one is based on artificial demand data drawn from random distributions. 
In the first case, we are given historical demand data for all local warehouses for a certain period of time and, for each day in this period, we try to fulfill all (historical) demand of that day in the order it occurred. 
In the second case, we assume that demands arrive according to a compound Poisson process. In this case, the mean of the Poisson arrival process and the mean and the variance of the order size of each arrival are given. We first draw the number of customers arriving from the Poisson distribution  (\ref{sec_appendix_poisson}) and then draw the order size for each customer. For the order sizes, the logarithmic distribution  (\ref{sec_appendix_logarithmic}) is used. This leads to a negative binomial distribution of the lead time demand, so we can use the fill rate formulas \eqref{eq:frord} in this case. When using real-world data, estimates for the mean and the variance of demand are available from the inventory planning system.

Finally, in the step \textbf{\textsc{S4}}, we check if the inventory position (i.e.,\ inventory on hand plus inventory on order minus backorders) at each warehouse is less or equal to the reorder point. If this is the case, we place as many backorders of size $Q_{i}$ at the preceding warehouse as necessary to raise the inventory position above $R_{i}$ again. If the preceding warehouse has enough inventory on hand, we fulfill the demand by scheduling a shipment as described in step \textbf{\textsc{S2}}.

\Cref{tab_SimuParameters} summarizes all parameters that can be configured in our simulation framework. It also shows the default settings that are used if not mentioned otherwise. 

We use a warm-up period, which should be a multiple of the longest transportation time in the network, to allow for a few order cycles to happen. We discard all measures obtained during this warm up period, except for the ones related to inventory, such as inventory on hand, inventory on order, and backorders.

\begin{table*}[htbp]
	\centering
		\begin{tabular}{l l l}
			Parameter & Input & Default value \\ \hline
			Transportation time &distribution& gamma\\
			Initial inventory & pieces & $R+1$\\		
			Warm-up period & time & -\\		
			Runtime& time & -\\
			Demand type&random or historical& -\\
			Order size&distribution&logarithmic
		\end{tabular}
	\caption{Parameters of simulation}
	\label{tab_SimuParameters}
\end{table*}

For each warehouse, we obtain the performance measures shown in \Cref{tab_SimuPerformance} as the result of the simulation. In addition, the order fill rate of each warehouse can be easily computed from these values as $\text{orders fulfilled}/\text{total orders}$. 

\begin{table*}[htbp]
	\centering
		\begin{tabular}{l L{10cm} }
			Performance measure & Description  \\ \hline
			Average inventory on hand&Average inventory on hand over all time periods\\
			Average inventory on order&Average of outstanding orders over all time periods\\
			Average backorders& Average of backorders over all time periods\\
			Total orders & Sum of incoming orders\\
			Orders fulfilled & Sum of incoming orders, that were fulfilled on the same day\\
			Wait time & Mean and variance of times a warehouse had to wait for replenishment orders from the time the order was made until it was shipped\\
		\end{tabular}
	\caption{Performance measures for each warehouse of the simulation}
	\label{tab_SimuPerformance}
\end{table*}
\section{Experimental results}\label{sec:exp}

In this section, we report on simulations that we have run to evaluate the accuracy of the different wait time approximations in our setting. 

\begin{figure}[htbp]
	\centering
		\includegraphics[width=0.8\textwidth]{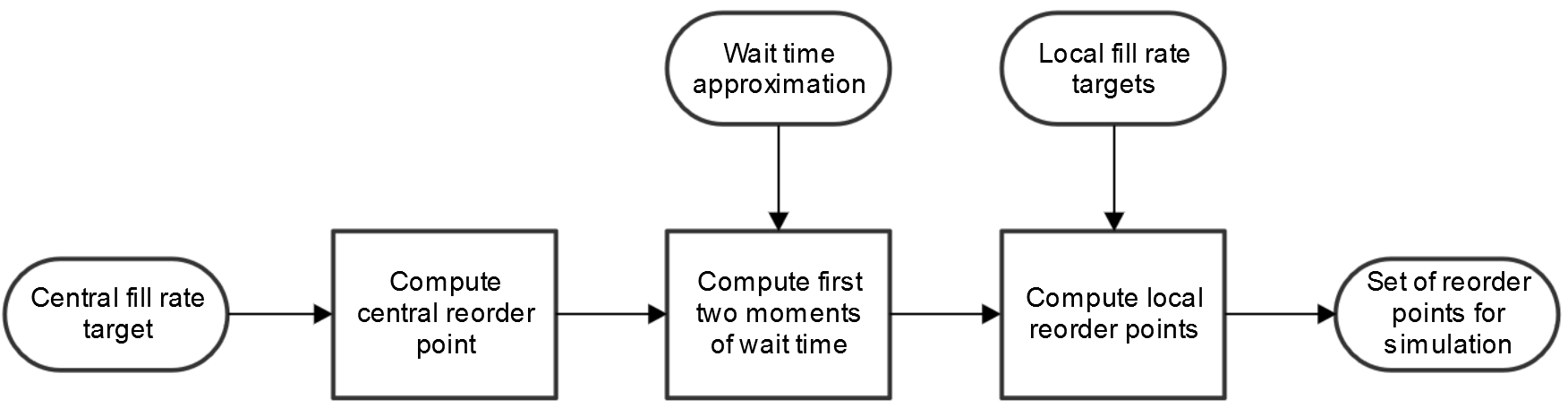}
	\caption{Process of calculating reorder points for the simulation given a prescribed central fill rate target, a wait time approximation and local fill rate targets}
	\label{fig_simu_set_rop}
\end{figure}

In our experiments, we considered a 2-level network and prescribed several central fill rate targets.
We then calculated the reorder points for all warehouses such that the given local fill rate targets are met. The local reorder points depend on the the wait time, so different wait time approximations will lead to different local reorder points. \Cref{fig_simu_set_rop} illustrates the process of calculating reorder points given a prescribed central fill rate, a wait time approximation and local fill rate targets. We want to emphasize that for the same prescribed central fill rate the resulting central reorder point is the same for all wait time approximations. The choice of approximation method only affects the local reorder points. Hence, 
the results are directly comparable.

The presentation of our experimental results is split in two sections. First, we try to establish a general understanding in which situation which approximation has the highest accuracy. For this purpose, we have set up experiments using an artificial network and random data drawn from certain distributions. In a second step, we try to verify our findings using the network and historical demand data of a real-world industrial supply chain. The second step is much more challenging than the first one, as the real-world demand data is ``dirty''. \cite{Wagner.2002} coined this term to describe that real-world demand data usually contains many irregularities, which often cannot be adequately represented by drawing random numbers based on a fixed probability distribution. The real-world demand data therefore can be seen as a more challenging environment and, consequently, a worse accuracy of the distribution-based wait time approximations should be expected. Nevertheless, any method that claims to have value in practical applications should be verified within a practical setting. To the best of our knowledge, we are the first to report on numerical tests of wait time approximations in a real-world setting with real-world demand data.

To evaluate the accuracy of the wait time approximations regarding the mean and standard deviation, we use two measures which we will define in the following for the \textsc{NB} approximation and which will be analogously used for all other approximations. 
Let $\mu_{NB}(t)$ and $\sigma_{NB}(t)$ be the computed estimators of the \textsc{NB} approximation for the mean and standard deviation of the wait time for test case $t$. Analogously, let $\mu_{SIMU}(t)$ and $\sigma_{SIMU}(t)$ be the simulated mean and standard deviation of the wait time for test case $t$. In \Cref{sec:wait_randon_data} we consider random demand data. Here, $\mu_{SIMU}$ and $\sigma_{SIMU}$ are the averages of the simulated values of all instances (in the respective group of instances) that were simulated.
We calculate the error and the absolute error of the mean of the wait time for the \textsc{NB} approximation for a set $T$ of test cases as 
\begin{align}
		\text{Error}=\frac{1}{|T|}\sum_{t\in T}(\mu_{NB}(t)-\mu_{SIMU}(t))\text{, and} \label{eq:error_wait_time}\\
		\text{Absolute Error}=\frac{1}{|T|}\sum_{t\in T}|\mu_{NB}(t)-\mu_{SIMU}(t)|\label{eq:absolute_error_wait_time}.
\end{align}
In the same fashion, we calculate error measures for the standard deviation and all other approximation methods. While \cref{eq:error_wait_time} conveys information about the average direction of the errors, \cref{eq:absolute_error_wait_time} conveys information about the average size of the errors.

\subsection{Random data}\label{sec:wait_randon_data}

Our objective in the experiments with random demand data is two-fold: First, we want to evaluate the accuracy of the given wait time approximations in different situations. In practice, however, managers do not care much about the wait time as a performance measure, they care much more if the resulting fill rate targets are fulfilled. As second objective, we thus also want to find out if using the wait time approximations to determine reorder points leads to small or big errors in the resulting fill rates. 
For this, we first compute the local reorder points for each test scenario and target fill rate using a wait time approximation, and then, in a second step, compare this with the actual fill rate observed for these reorder points in a simulation. 

\subsubsection{Set-up of experiments}

In order to compare the quality of the different wait time approximations for random demand data, we  consider a 2-level network, where warehouse $0$ is the central warehouse. 
The demand characteristics of the default case are shown in \Cref{tab_numericalsample}. 
\begin{table*}[htbp]
	\centering
		\begin{tabular}{C{1.2cm}| l r | r| r | r r | r }
			Ware- house& $\mu_i$& $\sigma_i^2$& $Q_i$&$\bar{\beta_i}$&$E[T_i]$&$\sigma(T_i)$&$p_i$\\ \hline
			0&-&-&500&-&60&30&0.5\\
			1&2&4&50&0.9&5&3&1\\
			2&3&6&50&0.9&5&3&1\\
			3&4&8&100&0.9&5&3&1\\
			4&5&10&100&0.9&5&3&1\\
			5&6&12&150&0.9&5&3&1\\
			6&7&14&150&0.9&5&3&1\\
			7&8&16&200&0.9&5&3&1\\
			8&9&18&200&0.9&5&3&1\\
		\end{tabular}
	\caption{Sample network characteristics}
	\label{tab_numericalsample}
\end{table*}
In this table, $\mu_i$ and $\sigma_i^2$ denote the mean and the variance of the demand for one day. From these values, one can easily derive the necessary parameters $\theta$ of the logarithmic order size and $\lambda$ of the Poisson arrivals as
\begin{align}
	\theta_i &=1-\frac{\mu_i}{\sigma_i^2} \text{~and}\\
	\lambda_i &=-\mu_i(1-\theta_i)\frac{log(\theta_i)}{\theta_i}~.
\end{align}
We have chosen to represent daily demand by its mean and variance, as this representation is probably more intuitive for practitioners. One can easily grasp the size and the variability of demand in the network. 

From this default case we derive our set of test cases by varying the parameters. \Cref{tab_SimuVariations} summarizes the test cases we consider. If the variation typ is multiplicative, indicated by `m' in \Cref{tab_SimuVariations}, we multiply the parameter by the variation value in the respective line of \Cref{tab_numericalsample} . If the variation typ is absolute `a', we replace it. A special case is the number $n$ of warehouses, where we vary the size of the network by considering the central warehouse $0$ and $n$ identical copies of warehouse $1$. Note, that we only vary one parameter at a time and keep all other parameters fixed to the values shown in \Cref{tab_numericalsample}.  Therefore, we have a total of 40 test cases, including the base scenario as shown in \Cref{tab_numericalsample}.
\begin{table*}[htbp]
	\centering
		\begin{tabular}{L{2cm}| l| l l}
			Parameter &\#Test cases& Variation &Variation typ \\ \hline
			$\mu_i$ &2 &0.25, 0.5 & m\\
			$\sigma_i^2$ &4 & 2, 4, 8, 16& m\\
			$Q_i, i>0$&5 &0.25, 0.5, 2, 4, 8 &m\\
			$Q_0$&5 &0.25, 0.5, 2, 4, 8 &m\\
			$\bar{\beta}_i$&4 &0.25, 0.5, 0.8, 0.95& a\\
			$T_0$&5 & 0.0625, 0.125, 0.25, 0.5, 2 &m\\
			$p_0$&3&2,4,8 &m\\
			$n$&10 &2,3,4,5,6,7,8,10,15,20& -\\
		\end{tabular}
	\caption{Parameter variations and variation type, multiplicative (m) or absolute (a), for the creation of the test cases}
	\label{tab_SimuVariations}
\end{table*}

Our objective is to test all wait time approximations over a wide range of central stocking quantities, from storing little at the central warehouse to storing a lot. We therefore prescribe different values for the central fill rate target. Based on the prescribed central fill rate, we then calculate the central reorder point that is necessary to fulfill the specified fill rate target as illustrated in \Cref{fig_simu_set_rop} to obtain sets of reorder points for the simulation. 

The different wait time approximations suggest different ways to compute the central fill rate. To obtain comparable results, we use the procedure described in \Cref{sec:model} to approximate the central lead time demand and then calculate the fill rate using \eqref{eq:frord}. 

In the simulations, we considered  a time horizon of 2000 days and used a warm-up period of 500 days. Complementary numerical experiments revealed that 100 independent runs of the simulation for each instance are sufficient to get reliable numerical results, so all reported simulation results are averages over 100 runs per instance. For further details on these tests we refer to \ref{sec_app_no_instances}.

In our initial experiment, we considered 5 scenarios where we prescribed a central fill rate of $20\%$, $40\%$, $70\%$, $90\%$  and $95\%$, respectively. We then calculated the central reorder point based on the prescribed central fill rate and equation \eqref{eq:frord} using a binary search and the procedure described in \Cref{sec:model}. Finally, we ran simulations to check if the intended central fill rate was actually obtained. As shown in \Cref{tab_SimuCentralRename}, especially in the high fill rate scenarios, the fill rates observed in the simulation are much lower than those anticipated.
\begin{table*}[htbp]
	\centering
		\begin{tabular}{c| c c}
		Prescribed central fr&Average simulated fr& Name of scenario\\ \hline
		20\%&10.86\%&low\\
		40\%&44.77\%&medium low\\
		70\%&56.06\%&/ \\
		90\%&63.59\%&/ \\
		95\%&67.80\%&medium high\\
		new&95.16\%&high
		\end{tabular}
	\caption{Prescribed central fill rate, average simulated fill rate and name of scenario}
	\label{tab_SimuCentralRename}
\end{table*}
The 5 initial scenarios only cover the low and medium range of central stocking quantities. 
As our objective is to cover a wide range of central stocking quantities in our tests, 
we manually adjusted the central reorder points for the base scenario and each test case in \Cref{tab_SimuVariations} (by trial and error with the help of the simulation) to create a new scenario that led to a high central fill rate of $95.16\%$ in the simulation. 

In the remaining experiments, we then only consider the scenarios $20\%$, $40\%$, $95\%$ and ``new'' shown in \Cref{tab_SimuCentralRename}, which have been renamed for clarity to align with the central fill rates actually observed in the simulations. Note that the ``high'' scenario is somewhat special: In this scenario, the central fill rate is so high that the chance an order has to wait at the central warehouse is really low. Hence, the effect of the wait time on the network is very small in this case. Nevertheless, we found it worthwhile to also consider this case for completeness.

For brevity, we refer to numbers obtained from the wait time approximations as \emph{computed} numbers and to the numbers observed in the simulation as
\emph{simulated} numbers in the following. 

\subsubsection{Results of the experiments}\label{sec_random_wait_results}

In this section we discuss the results of the experiments with random data described in the previous section. We compare simulated and computed wait time results. Based on this analysis we give recommendations when to use which approximation.  Then, we analyze if the local fill rate targets were satisfied and have a look at the inventory level for the different scenarios and approximations.

\paragraph{Wait time results}
\Cref{tab_waitapproxrand_wait} shows the average simulated and computed mean and standard deviation for each scenario and approximation. 
The \textsc{KKSL} approximation seems to perform best overall, while all other approximations seem to have great errors according to the simulated results.
However, we will observe later in this section that each approximation performs best for at least some scenarios and test cases. Hence, the aggregated results can only give a rough overview. Even more, some the aggregated results are strongly influenced by few individual cases where an approximation did perform very poor. For example, the standard deviation is badly overestimated by the \textsc{NB} approximation for a small number of parameter settings, although this approximation performs very good for many other cases. In the aggregate view, this is not properly represented.

\begin{table*}[htbp]
	\centering
		\begin{tabular}{c |r r| r r| r r| r r}
			&\multicolumn{2}{c|}{low}&\multicolumn{2}{c|}{medium low}&\multicolumn{2}{c|}{medium high}&\multicolumn{2}{c}{high}\\  
			&Mean&SD&Mean&SD&Mean&SD&Mean&SD \\  \hline
			Simulation&24.76&15.63&9.11&11.35&4.12&7.21&0.42&1.90\\ \hline
			\textsc{KKSL}&29.80&22.13&8.83&11.77&3.30&5.84&0.88&2.37\\
			\textsc{NB}&35.78&91.54&5.83&28.57&1.20&7.45&0.05&0.99\\
			\textsc{BF}&1.18&6.08&0.67&4.70&2.27&4.98&0.38&3.76\\
			\textsc{AXS}&25.15&9.60&5.35&6.96&0.73&1.63&0.11&1.05\\
		\end{tabular}
	\caption{Average wait time and standard deviation for different scenarios and wait time approximations in comparison to the simulated results}
	\label{tab_waitapproxrand_wait}
\end{table*}

In the following, we will analyze which approximation works best in which circumstances. When we refer to certain parameters being high or low, this is always with respect to our base setting shown in \Cref{tab_numericalsample}.

\begin{figure}[htbp]
	\centering
		\includegraphics[width=0.5\textwidth]{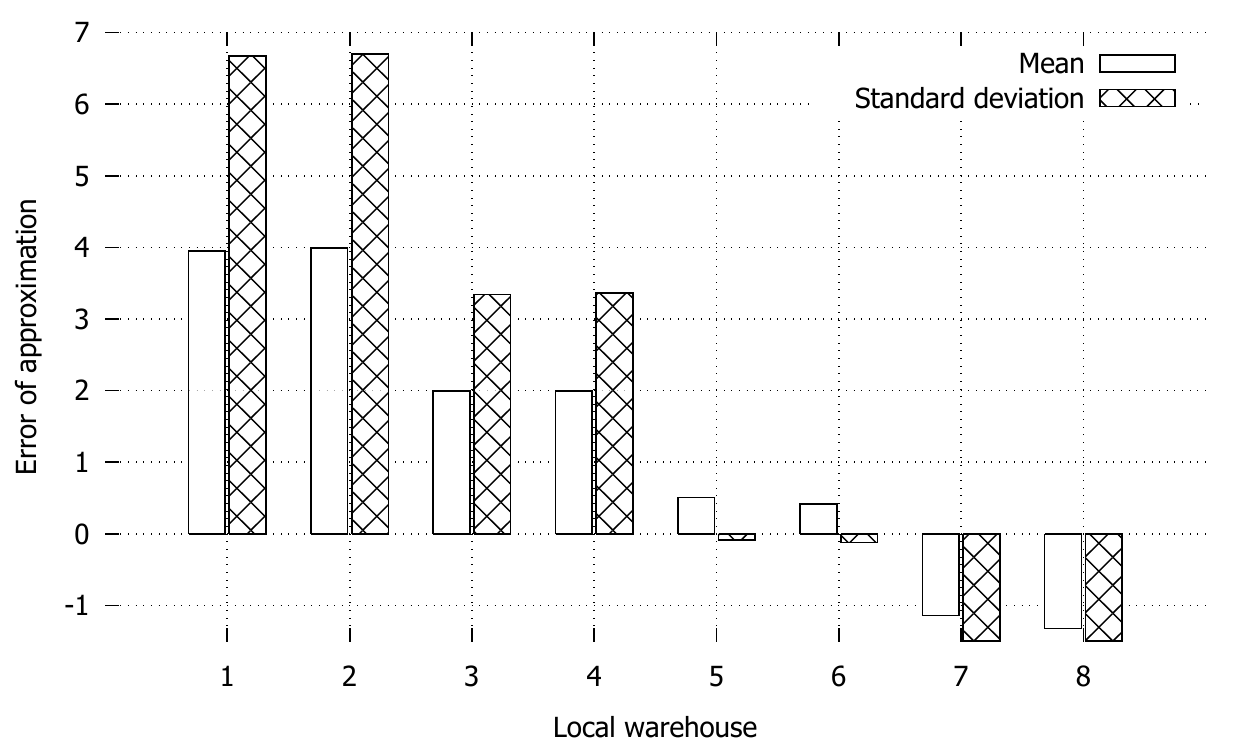}
	\caption{Errors of \textsc{KKSL} approximation for mean and standard deviation of wait time for the ``medium high'' scenario and the test set of base parametrization of \Cref{tab_numericalsample} for the different warehouses}
	\label{fig_kies_09_dev_base}
\end{figure}

\paragraph{Evaluation of the different approximations}
The \textsc{KKSL} approximation has good accuracy for a wide range of input parameters. It is rather accurate if the central fill rate is medium to high and if there are many local warehouses. It performs best if differences between local warehouses are small and if local order quantities are in a medium range, i.e., $Q_i/\mu_i \approx 20$ . Consider for example \Cref{fig_kies_09_dev_base}, which shows the errors for the mean and the standard deviation of the wait time for the``medium high'' scenario and on test instance, i.e., the base parametrization shown in \Cref{tab_numericalsample}. While the approximation performs good overall, the differences between local warehouses are large. The approximation is very good if the order quantity is neither too high nor too low, for example for warehouses 5 and 6, which have an  order quantity of 150. The accuracy is worst for warehouses with low demand and low order quantity. 

The \textsc{NB} approximation is very good for many input settings, especially at approximating the mean of the wait time. However, it badly overestimates the standard deviation  if $Q_i/\mu_i$ or if $Q_0/Q_i$ are large. On the other hand, we observed very good results with the \textsc{NB}  approximation when the central fill rate is medium to high and when the local warehouses are heterogeneous. 

In scenarios where the local order quantity and the central lead time are very low 
\textsc{BF} is a suitable approximation. In all other cases, it is not. We believe that the main reason for this is that it assumes a very smooth and nearly steady demand at the central warehouse. 
\cite{Berling.2014} state that this assumption is valid only for small $Q_i$. In the numerical analysis presented in their paper, increasing $Q_i$ causes larger errors. In our data, the ratio $Q_i/\mu_i$ is much larger than what Berling and Farvid considered. Moreover, our central lead times are much longer, so the system is not reset as frequently and the error of the linear approach supposedly adds up. 

By design, the \textsc{AXS} approximation computes only the average of the mean and the standard deviation of the wait time over all local warehouses. Consequently, it is only suitable if differences between local warehouses are not too large. Furthermore, it performs very good if the central fill rate is low or if the variance of the demand is high.
\bigskip

Next, we will have a closer look at the effect of the network size on the quality of the approximations. For this experiment, we consider a network with $n$ identical warehouses with the characteristics of warehouse 1 in \Cref{tab_numericalsample}. Note that this construction favors approximations that benefit from homogeneous networks, namely \textsc{KKSL} and \textsc{AXS}. We therefore focus on the trend and analyze how the accuracy of each individual approximation changes if we increase the number of local warehouses. 

\begin{figure}[htbp]
	\centering
	\subfloat[Mean]{
		\includegraphics[width=0.40\textwidth]{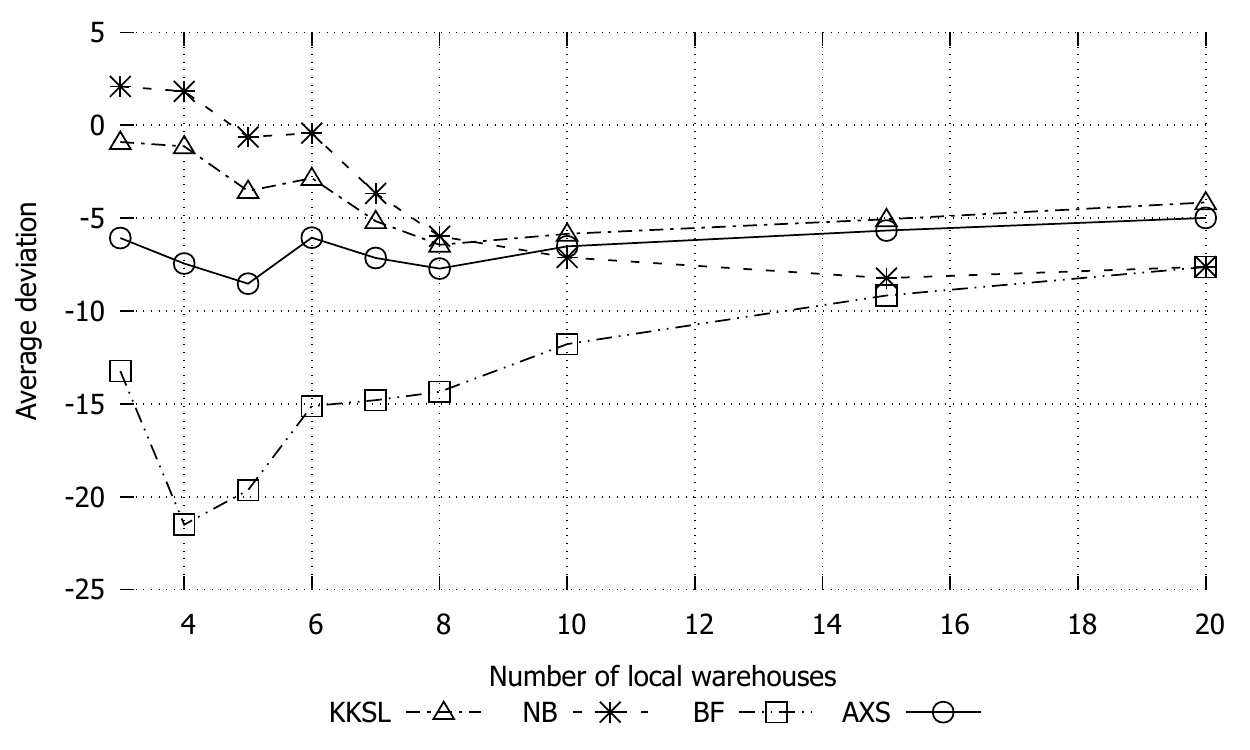}
		\label{fig_simu_node_04_mean}
	}
	\subfloat[Standard deviation]{
		\includegraphics[width=0.40\textwidth]{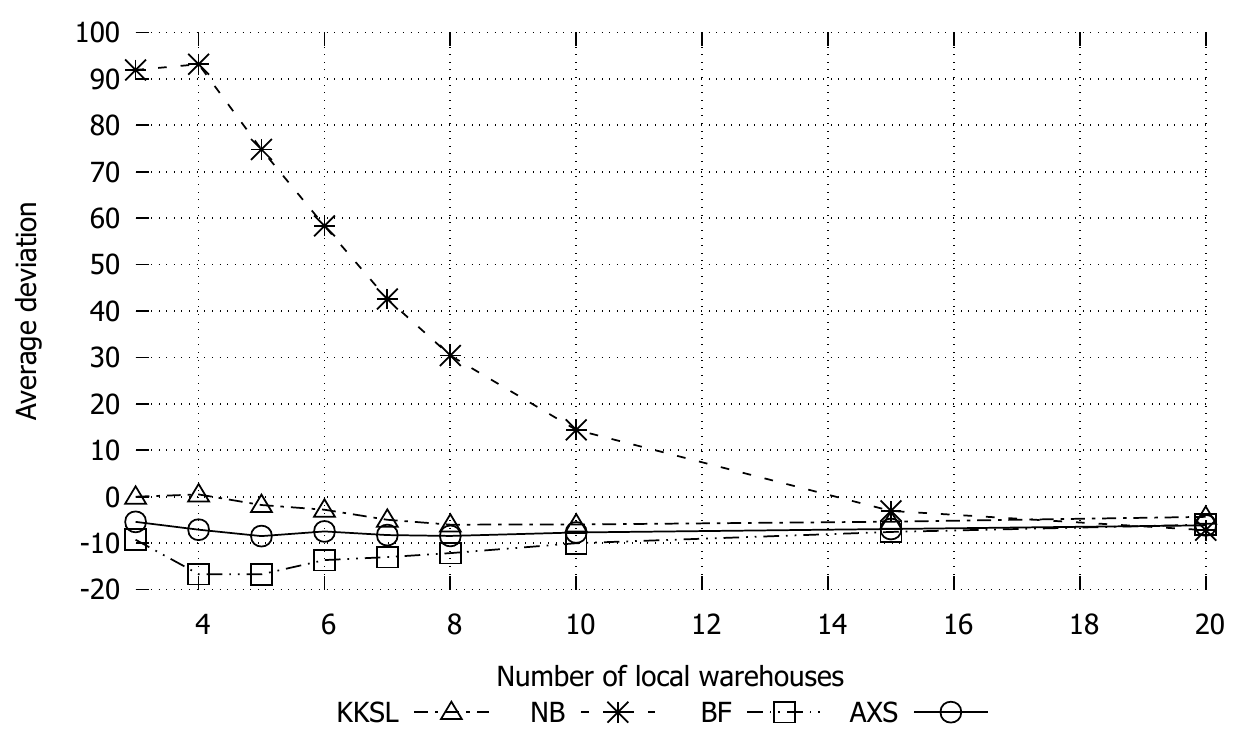}
		\label{fig_simu_node_04_sd}
	}		
	\caption{Errors of mean and standard deviation of the wait time for the different approximations and network sizes in the ``medium low'' scenario}
	\label{fig_simu_wait_node_04}
\end{figure}

\Cref{fig_simu_wait_node_04} shows the average error of the computed from the simulated mean and standard deviation of the wait time for the ``medium low'' scenario. For the mean of the wait time, we see no clear trend for $n<10$, but for $n\geq 10$ the accuracy of the approximation increases for all methods except \textsc{NB}, which levels off for $n>10$. 

For the standard deviation, we see an interesting behavior for the \textsc{NB} approximation: The approximation is bad for small networks, which is consistent with our previous analysis (as $\frac{Q_i}{\mu_i}=25$), but the quality dramatically improves with more local warehouses. This indicates that the mechanism that overestimates standard deviations for warehouses with those input parameters diminishes with increasing network size. The other methods show a slightly improving quality  with increasing network size.  This behavior is very similar for all other central fill rate scenarios.

\medskip

Finally, we look at the last test case, i.e., how do different wait time approximations perform if we change the local fill rate target ($\bar{\beta}_i$ in \Cref{tab_SimuVariations}). The local fill rate target has no effect on the accuracy of the approximations in our numerical experiments. This is hardly surprising as the local fill rate target only influences the local reorder point. Therefore only the timing of orders changes but neither the quantity nor the frequency of the ordering process. 

\paragraph{Local fill rates and inventory in network}
Ultimately, a planer does not care about wait time but is interested in whether the fill rate targets at the local warehouses are met or violated. Additionally, as a secondary objective, she or he might prefer fill rates to be not much higher than the targets, as higher fill rates require additional stock.

\Cref{tab_SimuFRlocalDevia} gives an overview on the average deviation of the local warehouses' fill rates from the fill rate targets. It shows the average of all test cases except the text ones regarding the network size $n$ and the fill rate targets $\bar{\beta}(i)$ (cp. \cref{tab_SimuVariations}). 
\begin{table*}[htbp]
	\centering
		\begin{tabular}{l| c c c c}
		Scenario&\textsc{KKSL} &\textsc{NB} &\textsc{BF} &\textsc{AXS} \\ \hline
		low&7.31\%&9.04\%&-46.64\%&-7.04\%\\
		medium low&3.36\%&5.10\%&-13.83\%&-8.39\%\\
		medium high&2.75\%&1.70\%&-2.15\%&-6.24\%\\
		high&5.13\%&2.51\%&5.19\%&0.22\%\\
		\end{tabular}
	\caption{Average deviations from fill rate targets for local warehouses, average of all test cases except the test cases for the network size $n$ and the fill rate targets $\bar{\beta(i)}$}
	\label{tab_SimuFRlocalDevia}
\end{table*}

\textsc{AXS} and \textsc{BF} on average violate the fill rate constraints considerably, except for the ``high'' scenario, where the quality of the wait time approximation is hardly important. This confirms our previous finding that these two wait time approximations are only suitable in some rather restrictive cases.  

For \textsc{KKSL} and \textsc{NB},  the fill rate constraints are -- on average -- fulfilled in most cases. However, for the ``low''\ and the ``medium low'' scenario the observed average fill rates are much higher than the fill rate targets, indicating a substantial overstock. 

\begin{figure}[htbp]
	\centering
	\subfloat[Reorder points]{
		\includegraphics[width=0.40\textwidth]{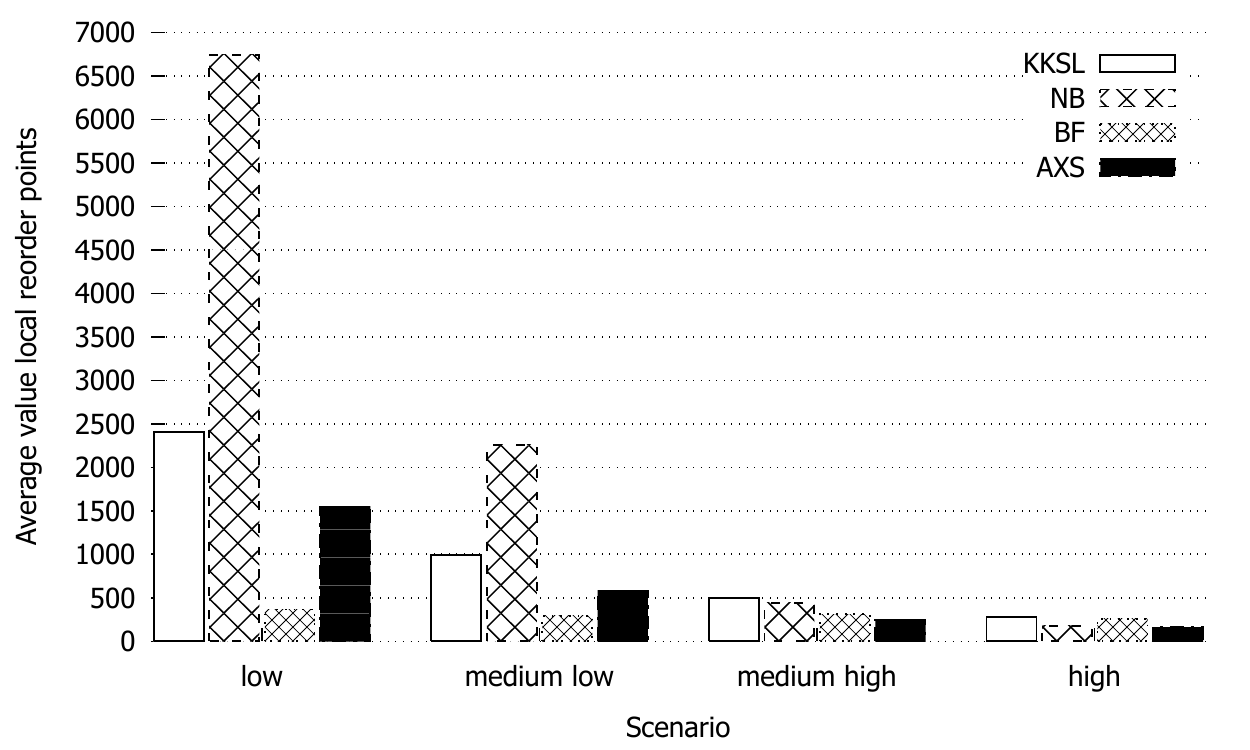}
		\label{fig_simu_stock_R}
	}
	\subfloat[Inventory levels]{
		\includegraphics[width=0.40\textwidth]{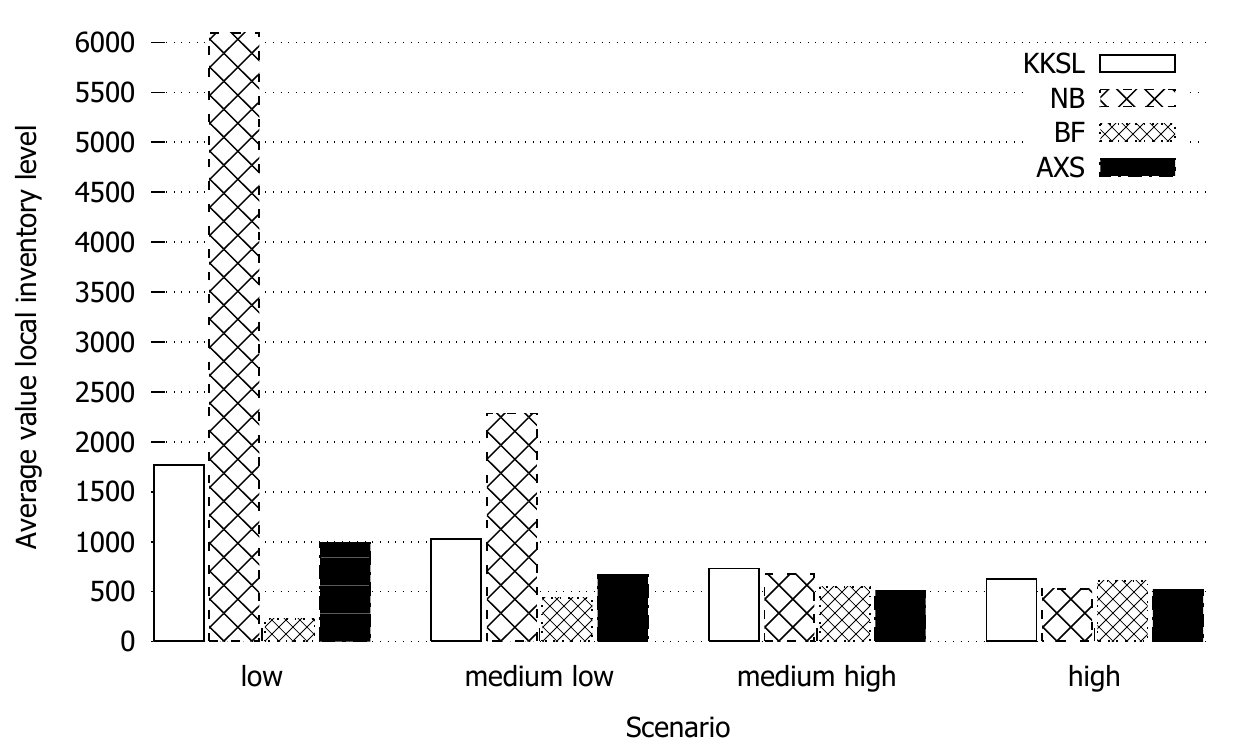}
		\label{fig_simu_stock_lev}
	}		
	\caption{Average value of reorder points and inventory levels for local warehouses for the different scenarios}
	\label{fig_simu_stock}
\end{figure}

\Cref{fig_simu_stock} shows the average value of the reorder points and inventory levels at local warehouses. As both bar charts show a very similar pattern, we discuss them together and more generically refer to stock instead of reorder points and inventory levels. 

Naturally, the local stock value is decreasing if more stock is kept at a central level. 

The stock situation for \textsc{AXS} and \textsc{BF} reflects the violation of fill rate targets: Whenever the violation is high, stock is low. Both the fill rate and the stock levels indicate that these two approximations underestimate the wait time.

The much higher stock for the \textsc{NB} for the ``low'' scenario is striking. Contrary, for the ``high'' scenario, results based on \textsc{NB} have the lowest stock. The higher the fill rate, the more the stock situation for \textsc{NB} is similar to the other approximations. The \textsc{KKSL} behaves more moderate than the \textsc{NB} approximation. We already found that the \textsc{NB} approximation dramatically overestimates the standard deviation of the wait time for some input parameter constellations. For these instances, way too much stock is kept. If we remove those instances from the analysis, the dramatic spike in the ``low'' scenario disappears. The \textsc{NB} approximation still overstocks for low central fill rates, but the accuracy for higher central fill rates is much better. One could assume that the underestimation of the average fill rate shown in \Cref{tab_SimuFRlocalDevia} is also highly influenced by this and that \textsc{NB} may violate the fill rate constraints if these instances are excluded from analysis. However, this is not the case: Although the average fill rate is lower, the fill rate targets are still met, see \Cref{tab_SimuFRlocalDevia_negbin}.

\begin{figure}[htbp]
	\centering
		\includegraphics[width=0.4\textwidth]{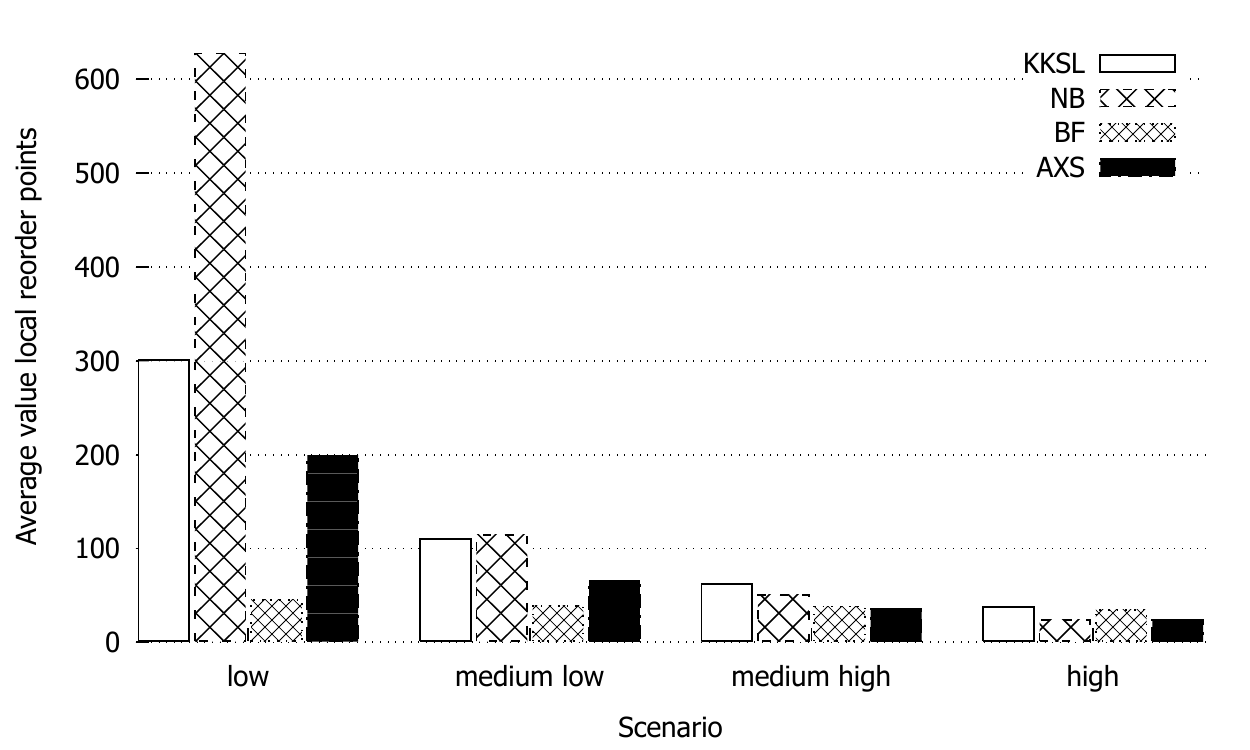}
	\caption{Average value of reorder points for local warehouses for the different scenarios excluding test cases not suitable for \textsc{NB}}
	\label{fig_simu_avgR_excluNB}
\end{figure}

\begin{table*}[htbp]
	\centering
		\begin{tabular}{l| c c c c}
		&low &medium low &medium high &high\\ \hline
		Deviation&8.46\%&3.04\%&1.15\%&2.42\%\\
		\end{tabular}
	\caption{Average deviation from fill rate target for local warehouses for \textsc{NB} approximation excluding test cases not suitable for \textsc{NB}}
	\label{tab_SimuFRlocalDevia_negbin}
\end{table*}

If we vary the local fill rate targets, the behavior of all approximations is not surprising: The higher the fill rate target, the more local stock is needed. The more accurate the approximation, the more precise the fill rate target is met and the less over- or understock, respectively, is observed.

\paragraph{Summary of the analysis based on random data}
The \textsc{KKSL} approximation seems to be a good default choice, unless local order quantities differ too much or the central fill rate is very low. A suitable alternative in many instances is the \textsc{NB} approximation, especially if the central fill rate is medium to high and if $\frac{Q_i}{\mu_i}<25$ or if there are many local warehouses. In these situations \textsc{NB} often outperforms the \textsc{KKSL} approximation. 

The \textsc{AXS} approximation should only be used if the characteristics of local warehouses are similar. If, additionally, the central fill rate is low or the variance of demand is very high, it is an excellent choice.

Only for very low central lead time and local order quantities, 
\textsc{BF} should be chosen as approximation.

\subsection{Real-world data}

The aim of the experiments presented in this section was to verify the findings for random data also in a real-world setting with real-world data. For this purpose, we considered a real-world distribution network with one central and 8 local warehouses and 445 parts from the total stock assortment in this network. For all parts, we were given fill rate targets, prices, order quantities and historical daily demand for all warehouses over a period of 1 year. The 445 parts were selected to make a small but representative sample of the total stocking assortment of over 80,000 parts. For the majority of the parts, the variance of the demand is very high compared to its mean.

In the simulation runs, the first 100 days were used as a warm-up period and results from those days were discarded. The lead times were generated from gamma distributions with given means and variances. For all scenarios considered here, the same random seed was chosen. 
In order to mimic a realistic real-world scenario, the reorder points were (re-)computed every 3 months based on (updated) demand distributions, order quantities, fill rate targets and prices. The demand distributions used as input in these computations were also updated every 3 months based on the
future demand forecast data from the productive inventory planning system.  

In the previous section, where we used artificial demand data, we followed an exploratory approach, manually searching for patterns in the data obtained in the computations and simulations. This was possible, because the size of the data set to be analyzed was sufficiently small. For the real-world data, 
this is not possible anymore. Therefore, we have chosen to take a different approach here: We state the key findings from the previous sections as hypotheses and try to confirm or falsify them here.

The mean and the variance of the transportation time from the central warehouse to each local warehouse is the same for all parts, but it differs between local warehouses.  
For each local warehouse, the mean is between 4 and 18 days and the variance is small, about $30\%$ of the mean. The lead time from the supplier to the central warehouse depends on the supplier and is therefore not the same for all parts. The mean is between 45 and 100 days. The variance is much higher, for some parts it is up to several factors larger than the mean. These characteristic values limit our analysis, so we cannot replicate all investigations from the previous section.

Also, the evaluation of the results contains an additional challenge. We constructed the simulation to mimic the behavior of a real-world distribution system, where demands change and reorder points are recalculated and adopted every 3 months. Every time the reorder points are updated, also the wait time approximation will change implicitly. The stock situation at the time of change, however, is determined by the results of the past. In the simulation, it will take some time until the updated reorder points will become visible in the actual stock and, thus, in the actual wait time. This behavior distorts the comparison of the approximation and the simulation, but this is something that would also happen if the approximations were implemented in a real-world system. To cope with this difficulty, we will compare the  mean and the standard deviation of wait time computed by the respective wait time approximation, averaged over all re-computations, to the mean and the standard deviation observed in the simulation over the entire time horizon excluding the warm-up period.
\bigskip

We start by looking at the central fill rate in order to verify the findings from \Cref{tab_SimuCentralRename}, namely that the simulated fill rate is much lower than the prescribed fill rate for high fill rate scenarios. The results shown in \Cref{tab_SimuCentral_realworld} clearly confirm this observation.  For the scenarios with a prescribed fill rate of 90\% and 95\%, the simulated fill rate is indeed much lower than those values.  For low prescribed fill rates, the simulated fill rate is higher than the prescribed value. Again we created an additional scenario by adjusting central reorder points by trial and error to obtain a high central fill rate scenario named ``new'', as similarly done for \Cref{tab_SimuCentralRename}.
\begin{table*}[htbp]
	\centering
		\begin{tabular}{l| c }
		Prescribed Fill Rate&Average Simulated Fill Rate\\ \hline
		20\%&35.22\%\\
		40\%&41.87\%\\
		70\%&56.55\%\\
		90\%&70.30\%\\
		95\%&76.93\%\\
		new&93.73\%
		\end{tabular}
	\caption{Prescribed central fill rate and average simulated value}
	\label{tab_SimuCentral_realworld}
\end{table*}

In total we have 18,396 test cases. A test case contains all results of our experiments for each combination of scenario, part number and warehouse. For example, the test case for scenario $20\%$, part ``a'' and warehouse ``1'' contains the results of the simulation as well as the a-priori calculated moments of the wait time for the four different approximations.

We use this big data set to test the following hypotheses, derived in the previous section for random demand data and specifically designed test cases. Note that our goal is not to confirm these hypotheses by any statistical means, but only to detect patterns in the data that either confirm or reject those findings.
\begin{itemize}[leftmargin=3.5em]
	\item[H1.] The \textsc{KKSL} approximation is suitable if the central fill rate is medium to high.
	\item[H2.] The \textsc{KKSL} approximation performs best if differences between local warehouses are small and if local order quantities are in a medium range.
	\item[H3.] The \textsc{NB} approximation overestimates the standard deviation significantly if \\$Q_i/\mu_i>25$, but it has a good accuracy if $Q_i/\mu_i\le 25$.
	\item[H4.] The \textsc{NB} approximation is good if the central fill rate is medium to high and when local warehouses are heterogeneous. 

	\item[H5.] The error of the \textsc{NB} approximation of the standard deviation reduces if the network becomes larger.
	\item[H6.] The \textsc{BF} approximation performs good if local order quantities and central lead time are small.
	\item[H7.] The \textsc{AXS} approximation performs good if the network is homogeneous, the central fill rate is low or the variance of demand is high.
\end{itemize}

\paragraph{Hypothesis H1: \textsc{KKSL} is suitable if the central fill rate is medium to high}

While hypothesis H1 explicitly states that \textsc{KKSL} is suitable for medium to high central fill rates, this also implies that it is not as suitable for low central fill rates. We will therefore analysis both situations, the accuracy of \textsc{KKSL} for medium to high central fill rates, corresponding to the 
4 scenarios 70\%, 90\%,, 95\% and ``new'' in \Cref{tab_SimuCentral_realworld},   and its accuracy for low central fill rates, corresponding to scenarios 20\%, 40\%.

\Cref{tab_Simu_realworld_H1} shows the ranking of the \textsc{KKSL} approximation compared to the other approximations. 
The columns ``mean" and ``standard deviation" show the respective percentages of the test cases, for which \textsc{KKSL} produced the best, at least the second best, or the worst of the 4 approximations for the value of the mean or the standard deviation of the wait time that was observed in the simulation. The column ``combined" shows the  
percentages of the test cases, where \textsc{KKSL} produced the best, at least second best, or worst values for both mean and standard deviation.

For medium to high central fill rates, 
\textsc{KKSL} is at least the second best approximation for the mean in about 50\% of the cases, but in  21.49\% of the cases it is also the worst method. The accuracy is better for the standard deviation. In more than 22\% percent of the cases, \textsc{KKSL} is best in approximating the mean and the standard deviation than any other method, as shown in the combined ranking. Only in less than 6\% of all cases it is the worst overall method. Looking at the results for low central fill rates,  we see that \textsc{KKSL} performs even better.
\begin{table*}[htbp]
	\centering
	\resizebox{\textwidth}{!}{%
		\begin{tabular}{l| c c c| c c c }
		&\multicolumn{3}{c|}{Medium to high central fill rate}&\multicolumn{3}{c}{Low central fill rate}\\
		Measurement&Mean&Standard deviation &Combined&Mean&Standard deviation &Combined\\ \hline  
		best &28.80\% &42.06\% &22.19\% &25.38\%&48.25\%&14.66\%\\  
		2nd or better&51.37\% &71.33\%&40.93\%&56.80\%&92.14\% &49.77\%\\
		worst &21.49\% &9.17\% &5.72\%&10.36\%&0.61\%&0.18\%
		\end{tabular}}
	\caption{Relative accuracy of the \textsc{KKSL} approximation for scenarios with medium to high fill rate relative to the other approximations}
	\label{tab_Simu_realworld_H1}
\end{table*}

\Cref{tab_Simu_realworld_H1_abs_dev} shows
the absolute error of the approximated values from the values observed in the simulation for the different methods, averaged over all test cases. Also these numbers support our observation that \textsc{KKSL} is even better for low central fill rates than for medium to high central fill rates. For medium to high central fill rates, the accuracy of \textsc{KKSL} for the mean is comparable to other methods, but its accuracy for the standard deviation is better.

\begin{table*}[htbp]
	\centering
		\begin{tabular}{l| c c| c  c }
			&\multicolumn{2}{c|}{Medium to high central fill rate}&\multicolumn{2}{c}{Low central fill rate}\\
			&Mean&Standard deviation&Mean&Standard deviation\\ \hline
			Simulation&5.22&4.27&16.72&10.35\\ \hline
			KKSL&4.43&4.94&8.85&7.11\\
			NB&4.30&11.93&9.08&43.92\\
			AXS&4.37&6.36&9.86&15.11\\
			BF&4.86&5.72&11.29&6.93\\
		\end{tabular}
	\caption{Average simulated values and absolute error of the approximations for test cases in different scenarios}
	\label{tab_Simu_realworld_H1_abs_dev}
\end{table*}

In order to see if certain approximation tend to generally over- or underestimate the mean or the standard deviation of the wait time, we also analyze the 
average positive or negative error of the approximations in \Cref{tab_Simu_realworld_H1_dev} instead of the the averaged absolute errors.
This additionally conveys information about the direction of the inaccuracies. For medium to high central fill rates, the mean is underestimated by all methods and \textsc{KKSL} seems to perform especially bad. For the standard deviation, there is a general overestimation tendency in all approximations and \textsc{KKSL} is very accurate.
\begin{table*}[htbp]
	\centering
		\begin{tabular}{l| c c| c  c }
			&\multicolumn{2}{c|}{Medium to high central fill rate}&\multicolumn{2}{c}{Low central fill rate}\\
			&Mean&Standard deviation&Mean&Standard deviation\\ \hline
			KKSL&-2.31&1.83&-1.56&4.61\\
			NB&-0.94&11.23&3.03&43.83\\
			AXS&-2.11&4.33&3.25&14.92\\
			BF&-2.01&2.73&-5.21&2.99\\
		\end{tabular}
	\caption{Error of the approximations in different scenarios}
	\label{tab_Simu_realworld_H1_dev}
\end{table*}

In total, \textsc{KKSL} seems to be a suitable choice for medium to high central fill rates. There are high inaccuracies, but these are not worse than with the other methods. For low central fill rates \textsc{KKSL} was very accurate compared to the other methods for the real-world data. We find that \textsc{KKSL} is suitable across all central fill rates and seems to be a good general choice relative to the other approximations.

\paragraph{Hypothesis H2: \textsc{KKSL} performs best if differences between local warehouses are small and if local order quantities are in a medium range}

In order to evaluate hypothesis H2, we have to more formally define what a small difference between local warehouses means. By talking about how different local warehouses are, we refer to the difference in order quantities as well as demand, i.e., the mean and variance of demand per time unit. We formally define how much the value $x_i$ of an individual local warehouse differs from the mean of all $x_i$ as 
\begin{align}
	\delta x_i:=\frac{|x_i-1/n\sum_{j=1}^nx_j|}{1/n\sum_{j=1}^nx_j},\label{eq_definition_difference}
\end{align}
where $x_i$ may be the order size $Q_i$, the mean demand $\mu_i$ or the variance of demand $\sigma^2_i$.

First, we analyze the case where the difference between warehouses is small only with respect to a measure and unrestricted with respect to the other measure, i.e., the three cases where $\delta Q_i\le 20\% \text{ for all }i=1,...,n$, $\delta \mu_i\le 20\%\text{ for all }i=1,...,n$ or $\delta \sigma^2_i\le 100\%\text{ for all }i=1,...,n$. The reason for the latter $100\%$ is that we have to use a larger difference to have enough test cases in this class.

Afterwards, we consider the case where the difference among the warehouses is small with respect to all three measures.
This combined evaluation of all three metrics, we have to use larger differences for $\delta Q_i$ and $\delta \mu_i$ as well. There we consider all test cases for which
 $\delta Q_i\le 40\% \cap \delta \mu_i\le 40\% \cap \delta \sigma^2_i\le 100\%\text{ for all }i=1,...,n$.

The second part of hypothesis H2 is that a local order quantity is in a medium range \textit{relative} to the size of demand. Insights from \Cref{sec_random_wait_results} show that the ratio of $Q_i/\mu_i$ should be in the area of $20$.

We start by evaluating test cases where differences in order quantity and demand of local warehouses are small as defined above compared to test cases where those differences are large. For the large differences, we only evaluate test cases where $\delta Q_i> 20\% \text{ for all }i=1,...,n$ for the order quantity and analogously for the other two metrics. Therefore, in those test cases the respective metric of each local warehouse differs substantially from its mean.

 \Cref{tab_Simu_realworld_H2_dev} summarizes the results of this comparison. Again, it shows the proportion of cases for which \textsc{KKSL} was best, second or better, or worst of all approximations for the mean, the standard deviation or both measures combined for different subsets of our test cases.  With these results, the first part of Hypothesis H2 cannot be confirmed. For our real-world data sets, the \textsc{KKSL} approximation does not perform significantly better or worse if differences between local warehouses are larger or smaller. There is even some indication that the relative accuracy of \textsc{KKSL} approximation does improve if differences are larger. The combined measure of mean and standard deviation is better for large differences of order quantity and mean demand while it is slightly worse for the variance of demand. Especially for the mean demand \textsc{KKSL} is the best approximation in $19.83\%$ of the cases for large differences compared to $12.65\%$ for small cases.

\begin{table*}[htbp]
	\centering
	\resizebox{\textwidth}{!}{%
		\begin{tabular}{l| c @{}c@{} c| c@{}  c@{} c|c@{}c@{}c }
		Diff.\ of \dots for&\multicolumn{3}{c|}{Mean}&\multicolumn{3}{c|}{Standard deviation}&\multicolumn{3}{c}{Combined}\\ 
		$\text{all }i=1,...,n$ &Best&2nd or better~ &Worst&Best&2nd or better~ &Worst&Best&2nd or better~ &Worst\\ \hline  
			$\delta Q_i$&&&&&&&&\\ 
			$\le 20\%$&33.95\%&54.76\%&25.19\%&38.42\%&71.19\%&3.77\%&17.47\%&39.78\%&1.88\%\\
			$>20\%$&26.68\%&52.93\%&16.62\%&45.01\%&79.37\%&6.71\%&20.03\%&44.51\%&4.18\%\\ \hline
			$\delta \mu_i$&&&&&&&&\\ 
			$\le 20\%$&24.38\%&45.37\%&34.88\%&42.28\%&70.06\%&7.41\%&12.65\%&32.10\%&5.86\%\\
			$>20\%$&27.73\%&53.34\%&17.42\%&44.16\%&78.44\%&6.29\%&19.83\%&44.12\%&3.83\%\\ \hline
			$\delta \sigma^2_i$&&&&&&&&\\ 
			$\le 100\%$&28.88\%&53.60\%&18.02\%&45.88\%&79.12\%&6.02\%&20.54\%&44.18\%&3.68\%\\
			$>100\%$&24.28\%&52.01\%&17.10\%&39.25\%&75.91\%&7.14\%&17.31\%&43.03\%&4.41\%\\ \hline
			$\delta Q_i\le 40\% \text{ and}$&&&&&&&&\\ 
			$\delta \mu_i\le 40\% \text{ and}$&26.25\%&45.80\%&32.41\%&37.66\%&67.45\%&7.22\%&13.39\%&29.92\%&4.72\%\\ 
			$\delta \sigma^2_i\le 100\%$&&&&&&&&\\ 
		\end{tabular}}
	\caption{Relative accuracy of \textsc{KKSL} approximation compared to other approximations for test cases with small and large differences between local warehouses, with difference defined as in \cref{eq_definition_difference}}
	\label{tab_Simu_realworld_H2_dev}
\end{table*}

\begin{figure}[htbp]
	\centering
	\includegraphics[width=0.60\textwidth]{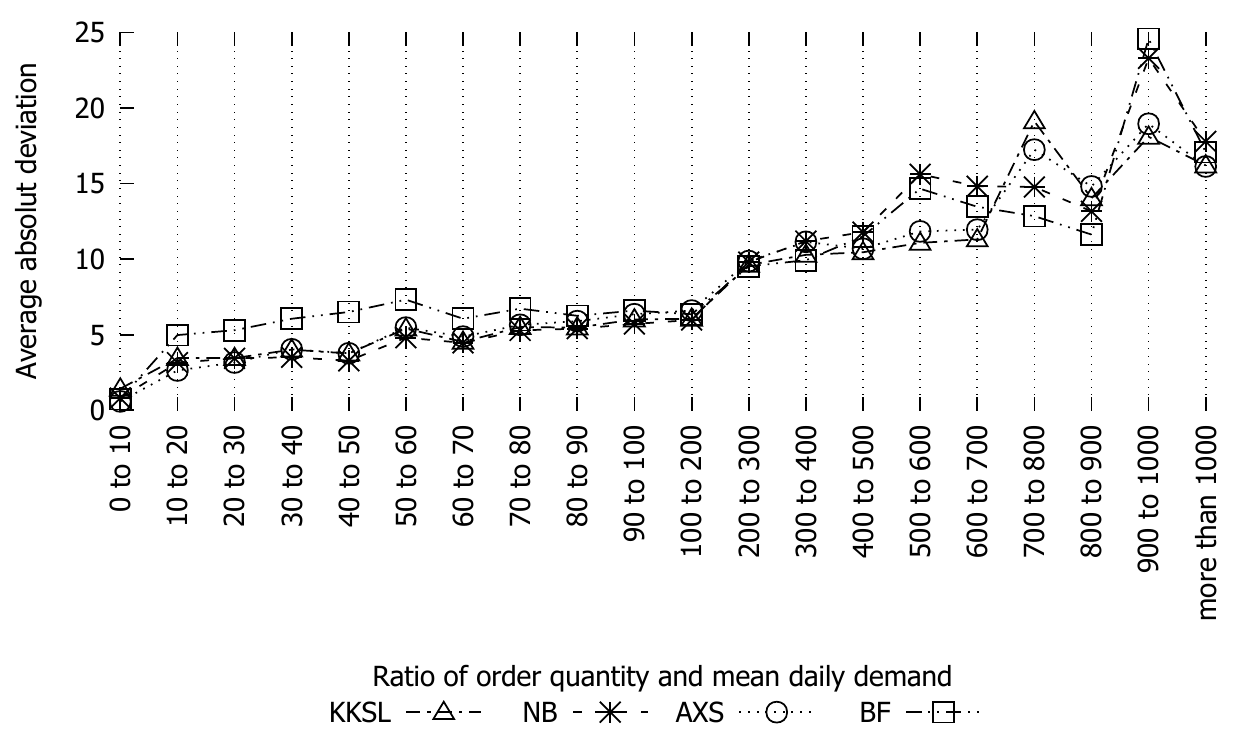}
	\caption{Absolute error of mean of the different approximations for different classes}
	\label{fig_realworld_wait_H2}
\end{figure}

\begin{figure}[htbp]
	\centering
	\includegraphics[width=0.6\textwidth]{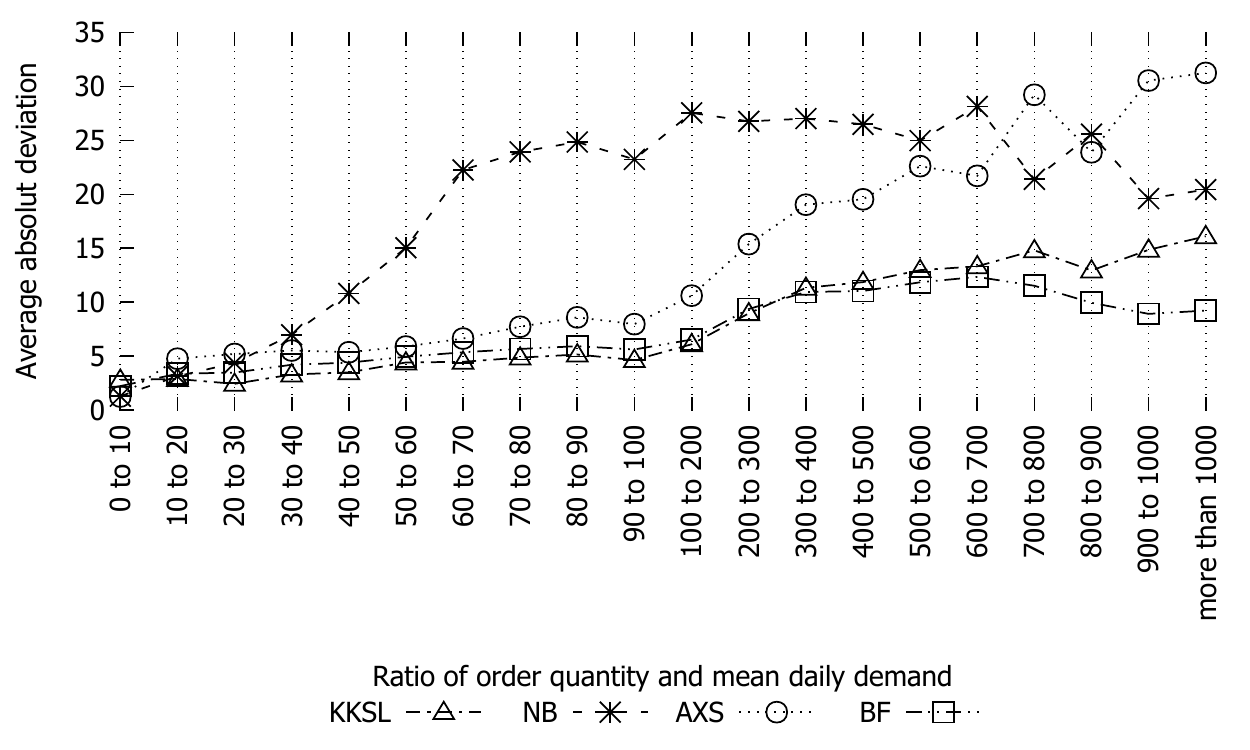}
	\caption{Absolute error of standard deviation of the different approximations for different classes}
	\label{fig_realworld_wait_H3}
\end{figure}

Regarding the order quantity, \Cref{fig_realworld_wait_H2} shows the average error from the simulated mean for different ratios of order quantity and mean demand. \Cref{fig_realworld_wait_H3} shows similar results for the standard deviation. Again, it cannot be concluded that the \textsc{KKSL} approximation  performs better or worse for order quantities in a medium range.

Based on the findings using the real-world data, we cannot confirm hypothesis H2. However, note that we could not intensively test the combination of only small differences between local warehouses and of order quantities in a medium range, because the number of test cases satisfying both conditions was too small.

\paragraph{Hypothesis H3: The \textsc{NB} approximation overestimates the standard deviation significantly if $Q_i/\mu_i>25$,  but it has a good accuracy if $Q_i/\mu_i\le 25$} 

To test hypothesis H3 we first compare all test cases for which $Q_i/\mu_i>25$ to all test cases for which $Q_i/\mu_i\le 25$. We want to emphasize that this overapproximation only occurs for the standard deviation and not for the mean. We look at the errors of the approximation, i.e., the difference between approximated standard deviation and simulated standard deviation, for these results. \Cref{tab_Simu_H3} shows the absolute average errir, separately for cases where $Q_i/\mu_i>25$ or $Q_i/\mu_i\leq 25$. Additionally, we state how many of the results fall in each category and what the average standard deviation of these results was. We also supply the results for the other approximations to confirm or reject that this is solely an issue for the \textsc{NB} approximation.
\begin{table*}[htbp]
	\centering
		\begin{tabular}{l c c| c c c c}
		&&Average sd &\multicolumn{4}{c}{Average absolute error}\\ 
		Case&\#test cases&in simulation&\textsc{NB} &\textsc{KKSL} &\textsc{AXS} &\textsc{BF}\\ \hline
		$>25$&15450&6.35&22.97&5.73&9.37&6.18\\
		$\le 25$&306&3.80&3.55&2.59&4.73&3.31\\
		\end{tabular}
	\caption{Absolute error of approximated standard deviation (sd) of wait time for test cases with different values of $Q_i/\mu_i$}
	\label{tab_Simu_H3}
\end{table*}

For the majority of our test cases we have $Q_i/\mu_i>25$. (Note that the number of test cases does not add up to the 18,396 sets in total, because we have to exclude the sets for the central warehouse, where wait time does not apply.  The results from the simulation with real-world data allows us to confirm hypothesis H3: The error is much larger for the standard deviation of the wait time if the \textsc{NB} approximation is used in the case $Q_i/\mu_i>25$. For $Q_i/\mu_i\le 25$, \textsc{NB} performs similar to the other approximations. Note that the other approximations also get worse if $Q_i/\mu_i$ is large, but they still perform much better than \textsc{NB} then.

A second interesting question is if the border is indeed rightly drawn at 25. \Cref{fig_realworld_wait_H3} shows the accuracy of the standard deviation predicted by the four approaches for different classes of $Q_i/\mu_i$. 
(Detailed results including the number of observations in each class are shown in \Cref{tab_Simu_realworld_H3_app} in the Appendix.) The error for the \textsc{NB} approximation gets large once $Q_i/\mu_i>30$. In all classes with $Q_i/\mu_i<30$, the quality of \textsc{NB} is comparable to the other approximations. So, 25 seems to be a good threshold value to decide whether \textsc{NB} should be used or not. Interestingly, the accuracy of \textsc{AXS} also becomes worse for increasing values of $Q_i/\mu_i$, although at a much higher threshold. For \textsc{KKSL} and \textsc{BF}, we only see a light deterioration in the approximation quality as $Q_i/\mu_i$ increases.

We now focus on test cases where $Q_i/\mu_i\le 25$ and try to establish if the \textsc{NB} approximation can be recommended in these circumstances. \Cref{tab_Simu_realworld_H3_dev} offers several insights: For the \textsc{NB} approximation, the approximation of the standard deviation is indeed much more accurate for 
$Q_i/\mu_i\le 25$ than it is in general. The accuracy of \textsc{KKSL} and \textsc{AXS} does also significantly benefit from excluding results with high local order quantity compared to mean demand. These three approximations all perform better than in the general case shown in
\Cref{tab_Simu_realworld_H1_dev}.  The \textsc{NB} approximation performs on average worse than \textsc{KKSL}. It is also better than \textsc{AXS} for the standard deviation, but worse than \textsc{AXS} for the mean. 

\begin{table*}[htbp]
	\centering
		\resizebox{\textwidth}{!}{%
		\begin{tabular}{l| c c| c  c | c c}
			&\multicolumn{2}{c|}{All senarios}&\multicolumn{2}{c|}{Medium to high central fill rate}&\multicolumn{2}{c}{Low central fill rate}\\
			&Mean&Standard deviation&Mean&Standard deviation&Mean&Standard deviation\\ \hline
			Simulation&5.01&3.80&3.23&2.87&8.57&5.66\\\hline
			KKSL&-1.72&-0.16&-1.47&-0.24&-2.23&0.00\\
			NB&-2.34&1.09&-2.12&-0.40&-2.79&4.07\\
			AXS&0.81&4.24&0.09&3.16&2.24&6.39\\
			BF&-4.26&0.73&-2.58&1.42&-7.62&-0.67\\
		\end{tabular}}
	\caption{Average simulated values and error of the approximations for different scenarios, for test cases with $Q_i/\mu_i\le 25$}
	\label{tab_Simu_realworld_H3_dev}
\end{table*}

A similar analysis was done also with the average absolute error instead of the average positive and negative error.
In order to compare also the absolute sizes of the approximation errors. 
The results of this analysis, which are shown in \Cref{tab_Simu_realworld_H3_abs_dev} in the Appendix, confirm the main observations from \Cref{tab_Simu_realworld_H3_dev}. The remarkably good accuracy of \textsc{AXS} in approximating the mean for medium to high central fill rate is due to cancel out effects.

Additionally, we repeat the analysis done in \Cref{tab_Simu_realworld_H1} for the \textsc{NB} approximation and the limited subset of results with $Q_i/\mu_i\le 25$. \Cref{tab_Simu_realworld_H3} shows the results. Overall, the \textsc{NB} approximation seems to be suitable if $Q_i/\mu_i\le 25$. As we have already pointed out, \textsc{KKSL} also benefits from this condition and therefore may be an overall better choice. \Cref{tab_Simu_realworld_H3_KKSL} replicates the results for \textsc{KKSL}. The share of cases where \textsc{KKSL} is best or second best is much lower than for \textsc{NB}. We report the results for \textsc{AXS} in \Cref{tab_Simu_realworld_H3_AXS} in the Appendix. \textsc{AXS}  does not perform better than \textsc{NB}. In particular, it has very little middle ground, especially for the mean: It is either the best approximation or the worst with a roughly even split. 

\begin{table*}[htbp]
	\centering
		\begin{tabular}{l| c c c| c c c| c c c  }
		&\multicolumn{3}{c|}{All scenarios}&\multicolumn{3}{c|}{Medium to high central fr}&\multicolumn{3}{c}{Low central fr}\\
		&Mean&sd &Comb.&Mean&sd &Comb.&Mean&sd &Comb.\\ \hline  
		Best&36.60\%&38.89\%&27.45\%&41.18\%&43.14\%&29.90\%&27.45\%&30.39\%&22.55\%\\
		2nd or better&61.11\%&60.78\%&48.69\%&61.76\%&70.10\%&55.88\%&59.80\%&42.16\%&34.31\%\\
		Worst&9.80\%&17.32\%&5.88\%&11.27\%&12.75\%&7.35\%&6.86\%&26.47\%&2.94\%\\
		\end{tabular}
	\caption{Relative accuracy of the \textsc{NB} approximation for the mean, the standard deviation (sd) and a combined (comb.) compared to other approximations.Evaluated  for different scenarios regarding the central fill rate (fr), only for test cases with $Q_i/\mu_i\le 25$ }
	\label{tab_Simu_realworld_H3}
\end{table*}

\begin{table*}[htbp]
	\centering
		\begin{tabular}{l| c c c| c c c| c c c  }
		&\multicolumn{3}{c|}{All scenarios}&\multicolumn{3}{c|}{Medium to high central fr}&\multicolumn{3}{c}{Low central fr}\\
		&Mean&sd &Comb.&Mean&sd &Comb.&Mean&sdn &Comb.\\ \hline  
		Best&14.38\%&25.16\%&8.17\%&14.22\%&25.98\%&8.33\%&14.71\%&23.53\%&7.84\%\\
		2nd or better&56.54\%&68.95\%&38.89\%&58.82\%&69.12\%&41.18\%&51.96\%&68.63\%&34.31\%\\
		Worst&11.11\%&14.71\%&9.80\%&13.24\%&15.69\%&11.27\%&6.86\%&12.75\%&6.86\%\\
		\end{tabular}
	\caption{Relative accuracy of the \textsc{KKSL} approximation for the mean, the standard deviation (sd) and a combined (comb.) compared to other approximations. Evaluated  for different scenarios regarding the central fill rate (fr), only for test cases with $Q_i/\mu_i\le 25$ }
	\label{tab_Simu_realworld_H3_KKSL}
\end{table*}

Overall, Hypothesis H3 is  supported by our simulations with real-world data.

\paragraph{Hypothesis H4: The \textsc{NB} approximation is good if the central fill rate is medium to high and when local warehouses are heterogeneous}

To evaluate hypothesis H4 we have to repeat the first part of the analysis of H2 and for the subset of scenarios with medium to high fill rates, i.e., scenarios 70\%, 90\%, 95\% and ``new'' in \Cref{tab_SimuCentral_realworld}. Based on the findings for hypothesis H3, we only consider cases with $Q_i/\mu_i\le 25$. Because of these restrictions, however, we can rely only on a relatively small number of test cases in the following. 

One could argue that by only considering cases with $Q_i/\mu_i\le 25$ we exclude all test cases where the \textsc{NB} approximation has a bad accuracy. We would then naturally get good results. We have chosen to do so to exclude the distorting effect of these test cases where the accuracy for the standard deviation of the \textsc{NB} approximation is really bad as established in the analysis of hypothesis H3. As we are looking at trends, i.e., does the accuracy improve if warehouses are more heterogeneous and if the fill rate is higher, we are still able to get general insights. However, we also repeated the analysis done here for all test cases. All findings reported in the following are supported by this data as well. 

The results are summarized in \Cref{tab_Simu_realworld_H4_dev}, which has the same structure as \Cref{tab_Simu_realworld_H2_dev}. There is indication that the \textsc{NB} approximation should be used if differences between local order quantities $\delta Q_i$ are large as it is more often more accuracte than the other approximations for mean and standard deviation of the wait time. 

For the differences in the mean of demand $\delta \mu_i$ the relative accuracy is worse for larger differences. 
For the variance of demand results are mixed and there is no clear indication if \textsc{NB} approximation profits from larger differences.

\begin{table*}[htbp]
	\centering
	\resizebox{\textwidth}{!}{%
		\begin{tabular}{l| c c c| c  c c|ccc }
		Diff.\ of \dots for &\multicolumn{3}{c|}{Mean}&\multicolumn{3}{c|}{Standard deviation}&\multicolumn{3}{c}{Combined}\\ 
		$\text{all }i=1,...,n$ &Best&2nd or better &Worst&Best&2nd or better &Worst&Best&2nd or better &Worst\\ \hline  
			$\delta Q_i$&&&&&&&&\\ 
			$\le 20\%$&37.50\%&64.29\%&19.64\%&41.96\%&69.64\%&17.86\%&25.00\%&58.04\%&13.39\%\\
			$>20\%$&45.65\%&58.70\%&1.09\%&44.57\%&70.65\%&6.52\%&35.87\%&53.26\%&0.00\%\\ \hline
			$\delta \mu_i$&&&&&&&&\\ 
			$\le 20\%$&50.00\%&78.13\%&7.81\%&42.19\%&81.25\%&9.38\%&31.25\%&70.31\%&6.25\%\\
			$>20\%$&37.14\%&54.29\%&12.86\%&43.57\%&65.00\%&14.29\%&29.29\%&49.29\%&7.86\%\\ \hline
			$\delta \sigma^2_i$&&&&&&&&\\ 
			$\le 100\%$&34.38\%&68.75\%&9.38\%&45.31\%&67.19\%&10.94\%&25.00\%&59.38\%&3.13\%\\
			$>100\%$&44.29\%&58.57\%&12.14\%&42.14\%&71.43\%&13.57\%&32.14\%&54.29\%&9.29\%\\ 
		\end{tabular}}
	\caption{Relative accuracy of \textsc{NB} approximation compared to other approximations for test cases with small and large differences between local warehouses, for medium to high central fill rate scenarios and $Q_i/\mu_i\le 25$ and with difference defined as in \cref{eq_definition_difference}}
	\label{tab_Simu_realworld_H4_dev}
\end{table*}

All in all, we have some supporting evidence that the \textsc{NB} approximation is good if the central fill rate is medium to high, when local warehouses have a large difference in order quantity while the difference between the mean demand should be rather small. This seems to hold also for the low central fill rate scenarios, but here also the share of cases where \textsc{NB} is the worst approximation increases, as shown in \Cref{tab_Simu_realworld_H4_dev_low}.
\begin{table*}[htbp]
	\centering
	\resizebox{\textwidth}{!}{%
		\begin{tabular}{l| c c c| c  c c|ccc }
		Diff.\ of \dots for  &\multicolumn{3}{c|}{Mean}&\multicolumn{3}{c|}{Standard deviation}&\multicolumn{3}{c}{Combined}\\ 
		$\text{all }i=1,...,n$ &Best&2nd or better &Worst&Best&2nd or better &Worst&Best&2nd or better &Worst\\ \hline  
			$\delta Q_i$&&&&&&&&\\ 
			$\le 20\%$&21.43\%&53.57\%&0.00\%&28.57\%&50.00\%&21.43\%&16.07\%&35.71\%&0.00\%\\
			$>20\%$&34.78\%&67.39\%&15.22\%&32.61\%&32.61\%&32.61\%&30.43\%&32.61\%&6.52\%\\ 
		\end{tabular}}
	\caption{Relative accuracy of \textsc{NB} approximation compared to other approximations for test cases with small and large differences between the order quantity of local warehouses, for low central fill rate scenarios and $Q_i/\mu_i\le 25$ }
	\label{tab_Simu_realworld_H4_dev_low}
\end{table*}

\paragraph{Hypothesis H5: The error of the \textsc{NB} approximation of the standard deviation reduces if the network becomes larger} 

This hypothesis is based on the findings in \Cref{fig_simu_wait_node_04}. There, we see a decreasing error of the \textsc{NB} approximation of the standard deviation. We have only limited means to test this in the real-world data, where we have 8 local warehouses. However, we do not have demand for all parts at all warehouses and, therefore, have results for which only a lower number of warehouses than 8 was considered.

\Cref{tab_Simu_realworld_H5} shows the error of the different approximations for all parts that ship to at most 3 local warehouses and for all parts that ship to all  8 local warehouses. \Cref{tab_Simu_realworld_H5_abs} in the Appendix shows the absolute error with similar findings. There is hardly a difference in the accuracy of the \textsc{NB} approximation between parts with at most 3 or exactly 8 local warehouses. We also checked if the share of cases where the \textsc{NB} approximation is the best or second best approximation changes (not presented here), but without finding any significant results. 

Therefore, we cannot confirm hypothesis H5. The only approximation that  seems to suffer from fewer warehouses in the real-world data is \textsc{AXS}. We suspect that this is caused by the fact that  this approximation considers only the mean across all warehouses and that variability is higher for fewer warehouses.

\begin{table*}[htbp]
	\centering
		\begin{tabular}{l| c c }
		\#Local warehouses&$8$&$\le 3$\\\hline
		Simulation&5.59&8.61\\\hline
		KKSL&2.35&2.98\\
		NB&21.23&21.20\\
		AXS&6.34&10.17\\
		BF&-4.13&-0.90\\
		\end{tabular}
	\caption{Mean of simulation and error of approximations for the standard deviation for results with different number of local warehouses}
	\label{tab_Simu_realworld_H5}
\end{table*}

\paragraph{Hypothesis H6: The \textsc{BF} approximation performs good if local order quantities and central lead time are small} 

Unfortunately, we cannot evaluate the two conditions of hypothesis H6 at the same time, as all test cases with a relatively low local order quantity also have a relatively low central lead time. As the minimal mean central lead time in our data set is about 45 days, it is questionable if we see effects at all. Nonetheless, we can evaluate the two conditions separately and compare mean central lead times of about 45 days to about 100 days.

\Cref{tab_Simu_realworld_H6_BF} shows the relative accuracy of the \textsc{BF} approximation for results with the two different lead times. The relative accuracy of the \textsc{BF} approximation is indeed better for the shorter central lead time.
\begin{table*}[htbp]
	\centering
		\begin{tabular}{l| c c c| c c c }
		Mean $L_0$:&\multicolumn{3}{c|}{$\approx 45$ days}&\multicolumn{3}{c}{$\approx 100$ days}\\
		Measurement&Mean&sd &Combined&Mean&sd &Combined\\ \hline  
		Best&23.78\%&29.49\%&12.99\%&11.11\%&35.42\%&9.38\%\\
		2nd or better&40.51\%&62.41\%&30.63\%&30.56\%&57.99\%&22.57\%\\
		Worst&41.75\%&9.63\%&7.65\%&47.57\%&9.38\%&7.99\%\\
		\end{tabular}
	\caption{Relative accuracy of the \textsc{BF} approximation for the mean and standard deviation (sd), comparison for high and low mean of central lead time compared to other approximations}
	\label{tab_Simu_realworld_H6_BF}
\end{table*}

To compare results for low and high local order quantities, we proceed as for hypothesis H3. In \Cref{tab_Simu_H3} we have already seen an improvement in the approximation of the standard deviation of \textsc{BF} for smaller local order quantities. Repeating the same analysis for the mean, find find that the quality of the \textsc{BF} approximation does not change. In fact, it stays remarkably constant. However, our limit of $Q_i/\mu_i\leq 25$ is already quite high compared to the ratios in the artificial data used in the previous section. Considering \Cref{fig_realworld_wait_H3} again, we see that the accuracy already deteriorates if the ratio is above 10, but it is quite good for the ratio being between 0 and 10. 

Thus, we have at least a good indication that \textsc{BF} profits from small order quantities and from smaller central lead times.

\paragraph{Hypothesis H7: The \textsc{AXS} approximation performs good if the network is homogeneous, the central fill rate is low or the variance of demand is high}

Hypothesis H8 consists of three conditions that we analyze separately. The statement concerning homogeneous networks is closely related to hypothesis H2. We therefore repeat the analysis that we did for \Cref{tab_Simu_realworld_H2_dev} in \Cref{tab_Simu_realworld_H8_homo} for the \textsc{AXS} approximation. We see that the accuracy of \textsc{AXS} relative to the other approximations does not improve for the real-world data. In fact, and in contrast to to the hypothesis, the relative accuracy of \textsc{AXS}  decreases for more homogeneous networks. This indicates that the other approximations, especially \textsc{KKSL}, benefit even more from similar warehouses.

\begin{table*}[htbp]
	\centering
	\resizebox{\textwidth}{!}{%
		\begin{tabular}{l| c c c| c  c c|ccc }
		Diff.\ of \dots for &\multicolumn{3}{c|}{Mean}&\multicolumn{3}{c|}{Standard deviation}&\multicolumn{3}{c}{Combined}\\ 
		$\text{all }i=1,...,n$ &Best&2nd or better &Worst&Best&2nd or better &Worst&Best&2nd or better &Worst\\ \hline  
			$\delta Q_i$&&&&&&&&\\ 
			$\le 20\%$&21.19\%&53.63\%&18.08\%&13.32\%&28.06\%&31.50\%&6.69\%&17.42\%&7.06\%\\
			$>20\%$&21.28\%&48.72\%&24.33\%&18.71\%&43.28\%&8.92\%&7.05\%&26.02\%&4.44\%\\ \hline
			$\delta \mu_i$&&&&&&&&\\ 
			$\le 20\%$&18.21\%&44.44\%&21.30\%&11.42\%&29.01\%&19.44\%&3.09\%&12.35\%&5.25\%\\
			$>20\%$&21.33\%&49.49\%&23.53\%&18.12\%&41.49\%&11.81\%&7.08\%&25.12\%&4.78\%\\ \hline
			$\delta \sigma^2_i$&&&&&&&&\\ 
			$\le 100\%$&21.16\%&50.98\%&22.21\%&17.96\%&41.45\%&10.34\%&6.74\%&25.06\%&3.15\%\\
			$>100\%$&21.58\%&44.97\%&27.01\%&18.06\%&40.61\%&16.48\%&7.73\%&24.31\%&9.34\%\\\hline
			$\delta Q_i\le 40\%\text{ and}$&&&&&&&&\\ 
			$\delta \mu_i\le 40\%\text{ and}$&20.47\%&49.74\%&20.87\%&10.76\%&28.48\%&23.36\%&3.02\%&12.07\%&3.94\%\\
			$\delta \sigma^2_i\le 100\%$&&&&&&&&\\ 
		\end{tabular}}
	\caption{Relative accuracy of \textsc{AXS} approximation compared to other approximations for test cases with small and large differences between local warehouses, with difference defined as in \cref{eq_definition_difference}}
	\label{tab_Simu_realworld_H8_homo}
\end{table*}

A comparison of the different approximations for a low central fill rate is presented in \Cref{tab_Simu_realworld_H3_dev,tab_Simu_realworld_H3_abs_dev} in the Appendix.  These results show that \textsc{AXS} does not perform significantly better than other approximations if central fill rates are low.

\Cref{tab_Simu_realworld_H8_dev_var} shows the average error of the approximation from the simulated values for different ratios of variance to mean of demand. (\Cref{tab_Simu_realworld_H8_abs_dev_var} in the appendix shows the absolute errors). It seems that \textsc{AXS} performs best if the ratio $\sigma^2/\mu$ is between 1 and 5 and its accuracy decreases for higher variance.

\begin{table*}[htbp]
	\centering
		\begin{tabular}{l| c c| c  c | c c|cc}
			&\multicolumn{2}{c|}{$\sigma^2/\mu<1$}&\multicolumn{2}{c|}{$\sigma^2/\mu<5$}&\multicolumn{2}{c|}{$\sigma^2/\mu\ge 1$}&\multicolumn{2}{c}{$\sigma^2/\mu\ge 5$}\\
			&Mean&sd&Mean&sd&Mean&sd&Mean&sd\\ \hline
			Simulation&7.58&5.81&8.78&6.18&9.20&6.35&10.47&6.93\\\hline
			KKSL&-0.44&3.65&-1.73&3.08&-2.23&2.67&-3.77&1.06\\
			NB&2.16&24.52&0.78&23.01&0.20&21.86&-1.67&17.38\\
			AXS&1.37&8.77&0.05&8.23&-0.49&7.77&-2.25&6.00\\
			BF&-0.67&3.87&-2.45&3.20&-3.32&2.71&-6.34&0.88\\
		\end{tabular}
	\caption{Average simulated values and errors of the approximations for the mean and standard deviation (sd), for test cases with different values of $\sigma^2/\mu$}
	\label{tab_Simu_realworld_H8_dev_var}
\end{table*}

We did not find evidence to confirm hypothesis H8.

\paragraph{Summary of the analysis based on real-world data}

For our real-world data, the results concerning the accuracy of the wait time approximations compared to the simulation are sobering. We often see large errors of all considered approximations. 
We have (at least some) supporting evidence for hypothesis H1, H3, H4, and H6,  but we can neither clearly confirm nor clearly reject the other hypotheses H2, H5, and H8.

\section{Summary and Conclusion}
In this paper, we presented the results of our extensive numerical experiments to analyze and compare the quality of different wait time approximations proposed in the literature. In our study, we considered both random as well as real-world demand data. 

Our experiments show that there is no generally ``best'' approximation, none of the wait time approximations has a better accuracy than the others in all situations.  In fact, it depends heavily on the characteristics of demand and the network which approximation performs best. 
In our experiments, we also observed rather large errors of the wait time approximations in some specific situations. This clearly shows room for improved models, at least for these situations, and the need for further research in this area. 

Nevertheless, we were able to derive a few simple guidelines that describe which approximation is likely to perform accurate in which situation:
The \textsc{BF} approximation should only be used if the local order quantities and central lead times are low. 
The \textsc{AXS} approximation is suitable for homogeneous networks, i.e., networks in which local warehouses are very similar with regards to demand structures and order quantities.
The \textsc{KKSL} and \textsc{NB} are generally good choices in all other circumstances. For the \textsc{NB} approximation, however,  the ratio $Q_i/\mu_i$ is critical for the quality of the approximation of the standard deviation of wait time. \textsc{NB}  should be used only if this ratio is smaller than 25.

In order to decide which approximation to use in a real-world application, we recommend to pick one or two likely candidates with the help of these guidelines and then perform additional simulations for the specific demands and the network arising in this application.

\addcontentsline{toc}{section}{References}
\bibliography{quellen2}
\newpage
\appendix
\section{Number of instances needed}\label{sec_app_no_instances}

We determined how many instances of each variation need to be run in the simulation to obtain dependable results. For this, we ran a number of tests for all variations of one scenario. The results, which are summarized in \Cref{tab_SimuInstances} for the mean simulated fill rate as an example,  are quite stable. Even if we look at mean simulated fill rates for an individual instance, we get only small deviations after 100 instances. We therefore decided that 100 instances are sufficient. We repeated the analysis with the mean and the standard deviation of the wait time, and also regarding these performance measures 100 instances seemed to be sufficient to get reliable results.
\begin{table*}[htbp]
	\centering
		\begin{tabular}{L{2cm}| l}
			\# instances &Mean fill rate \\ \hline
			10&87.534\% \\ 
			25&87.421\% \\ 
			50&87.299\% \\ 
			100&87.393\% \\ 
			200&87.350\% \\ 
			500&87.414\% \\
			1000&87.391\% \\
		\end{tabular}
	\caption{Mean simulated fill rates over all variations and instances for local warehouses}
	\label{tab_SimuInstances}
\end{table*}

\section{Supplementary tables}
\label{sec_appendix_data}

\begin{table}[htbp]
	\centering
		\begin{tabular}{ l c c| c c c c}
			Class&No. observations&Mean simulated&NB&KKSL&AXS&BF\\
			&&Standard deviation&&&&\\\hline
			0 to 10&30&1.40&1.30&2.82&1.27&2.23\\
			10 to 20&156&3.63&3.16&2.89&4.77&3.38\\
			20 to 30&258&4.48&4.38&2.45&5.21&3.49\\
			30 to 40&348&4.80&7.00&3.32&5.53&4.22\\
			40 to 50&390&6.14&10.83&3.50&5.37&4.41\\
			50 to 60&804&7.27&15.06&4.43&5.91&4.96\\
			60 to 70&4416&7.26&22.26&4.49&6.65&5.37\\
			70 to 80&2772&7.59&23.94&4.87&7.74&5.68\\
			80 to 90&1902&6.83&24.85&5.17&8.60&5.94\\
			90 to 100&306&7.40&23.24&4.65&7.98&5.63\\
			100 to 200&2580&5.98&27.55&6.12&10.62&6.55\\
			200 to 300&528&3.66&26.77&8.98&15.40&9.37\\
			300 to 400&312&2.25&27.03&11.34&19.05&10.96\\
			400 to 500&210&2.72&26.49&11.87&19.55&11.05\\
			500 to 600&132&0.80&24.98&12.98&22.61&11.86\\
			600 to 700&126&1.53&28.15&13.34&21.71&12.35\\
			700 to 800&54&0.58&21.39&14.81&29.23&11.53\\
			800 to 900&90&0.65&25.57&12.95&23.90&9.95\\
			900 to 1000&30&2.79&19.61&14.87&30.54&8.92\\
			$>$1000&312&0.41&20.44&16.11&31.26&9.23\\
		\end{tabular}
		\caption{Absolute error of the standard deviation for test cases with different ratios of order quantity and mean daily demand, Analysis of H3}
	\label{tab_Simu_realworld_H3_app}
\end{table}

\begin{table*}[htbp]
	\centering
		\begin{tabular}{l| c c c| c c c| c c c  }
		&\multicolumn{3}{c|}{All scenarios}&\multicolumn{3}{c|}{Medium to high central fr}&\multicolumn{3}{c}{Low central fr}\\
		&Mean&sd &Comb.&Mean&sd &Comb.&Mean&sd &Comb.\\ \hline  
		Best&38.24\%&19.61\%&17.32\%&35.29\%&20.59\%&18.14\%&44.12\%&17.65\%&15.69\%\\
		2nd or better&43.46\%&30.39\%&28.43\%&42.65\%&31.86\%&29.41\%&45.10\%&27.45\%&26.47\%\\
		Worst&48.37\%&51.63\%&39.54\%&47.55\%&51.47\%&36.76\%&50.00\%&51.96\%&45.10\%\\
		\end{tabular}
	\caption{Relative accuracy of the \textsc{AXS} approximation regarding the mean, standard deviation (sd) and combined (comb.) for different central fill rates (fr) and test cases with $Q_i/\mu_i\le 25$ }
	\label{tab_Simu_realworld_H3_AXS}
\end{table*}

\begin{table*}[htbp]
	\centering
		\begin{tabular}{l| c c| c  c | c c}
			&\multicolumn{2}{c|}{All senarios}&\multicolumn{2}{c|}{Medium to high central fill rate}&\multicolumn{2}{c}{Low central fill rate}\\
			&Mean&sd&Mean&sd&Mean&sd\\ \hline
			Simulation&5.01&3.80&3.23&2.87&8.57&5.66\\\hline
			KKSL&3.18&2.59&2.49&2.32&4.56&3.14\\
			NB&3.04&3.55&2.41&2.09&4.31&6.46\\
			AXS&2.51&4.73&1.81&3.85&3.90&6.49\\
			BF&4.70&3.31&3.13&3.44&7.84&3.03\\
		\end{tabular}
	\caption{Average simulated values and absolute errors of mean and standard deviation (sd) of the approximations for different scenarios, for test cases with $Q_i/\mu_i\le 25$}
	\label{tab_Simu_realworld_H3_abs_dev}
\end{table*}

\begin{table*}[htbp]
	\centering
		\begin{tabular}{l| c c }
		\#Local warehouses&$8$&$\le 3$\\\hline
		Simulation&5.59&8.61\\\hline
		KKSL&5.00&5.76\\
		NB&21.67&21.73\\
		AXS&7.46&11.78\\
		BF&5.61&5.48\\
		\end{tabular}
	\caption{Mean of simulation and absolute error of the approximations for the standard deviation for test cases with different number of local warehouses}
	\label{tab_Simu_realworld_H5_abs}
\end{table*}

\begin{table*}[htbp]
	\centering
		\begin{tabular}{l| c c| c  c | c c|cc}
			&\multicolumn{2}{c|}{$\sigma^2/\mu<1$}&\multicolumn{2}{c|}{$\sigma^2/\mu<5$}&\multicolumn{2}{c|}{$\sigma^2/\mu\ge 1$}&\multicolumn{2}{c}{$\sigma^2/\mu\ge 5$}\\
			&Mean&sd&Mean&sd&Mean&sd&Mean&sd\\ \hline
		Simulation&7.58&5.81&8.78&6.18&9.20&6.35&10.47&6.93\\\hline
		KKSL&5.24&5.83&5.78&5.80&5.97&5.65&6.57&4.94\\
		NB&5.77&24.83&5.85&23.44&5.90&22.37&6.09&18.21\\
		AXS&5.94&9.78&6.14&9.55&6.23&9.23&6.51&7.88\\
		BF&6.40&6.33&6.82&6.22&7.06&6.11&7.93&5.67\\
		\end{tabular}
	\caption{Average simulated values and absolute error of mean and standard deviation (sd) of the approximations, for test cases with different values of $\sigma^2/\mu$}
	\label{tab_Simu_realworld_H8_abs_dev_var}
\end{table*}

\newpage

\section{Distributions}
\label{sec_appendix_distributions}
In this section we define the required distributions in alphabetical order, as there are sometimes ambiguities in the definitions. We refer to the random variable as $X$, the probability mass function as $f_X(x)$  and the distribution function as $F_X(x)$.

\subsection{Compound Poisson}
\label{sec_appendix_compoundpoisson}
Let $N$ be a Poisson distributed random variable and $X_1, X_2, X_3,\dots$ i.i.d random variables that are also independent of $N$. Then
\begin{align}
	Y=\sum_{i=1}^{N}X_i
\end{align}
is compound Poisson distributed.

\subsection{Logarithmic distribution}
\label{sec_appendix_logarithmic}
Let $\theta \in (0,1)$ be the shape parameter of the logarithmic distribution. The probability mass function is
\begin{align}
	f_X(x)=\frac{-\theta^x}{x \ln(1-\theta)} \quad\text{for } x=1,2,3,\dots
\end{align}
and the probability distribution function is
\begin{align}
	F_X(x)=\frac{-1}{\ln(1-\theta)}\sum_{i=1}^x\frac{\theta^i}{i} \quad\text{for } x=1,2,3,\dots
\end{align}

\subsection{Negative binomial distribution}
\label{sec_appendix_negbin}
The density function of the negative binomial distribution with parameters $n>0$ and $0< p < 1$ is
\begin{align}
	f_X(x)=\frac{\Gamma(n+x)}{\Gamma(n)x!}p^n(1-p)^x\quad \text{for } x=0,1,2,\dots
\end{align}
where $\Gamma$ is the gamma function.
The probability distribution function is
\begin{align}
	F_X(x)=1-I_p(k+1,r) \quad\text{for } x=0,1,2,\dots
\end{align}

\subsection{Poisson distribution}
\label{sec_appendix_poisson}
Let $\lambda>0$ be the mean of the Poisson distribution. The probability mass function is
\begin{align}
	f_X(x)=\frac{e^{-\lambda}\lambda^x}{x!} \quad\text{for } x=0,1,2,\dots
\end{align}
and the probability distribution function is
\begin{align}
	F_X(x)=e^{-\lambda} \sum_{j=0}^x\frac{\lambda^j}{j!}  \quad\text{for } x=0,1,2,..
\end{align}

\end{document}